\documentstyle[epsf,seceq,preprint]{ptptex}

\def\be{\begin{equation}}
\def\ee{\end{equation}}
\def\bea{\begin{eqnarray}}
\def\eea{\end{eqnarray}}

\def\simm#1{\mathop{\vtop{\ialign{##\crcr
        $\hfil\displaystyle{#1}\hfil$\crcr\noalign{\kern0.5pt\nointerlineskip}
        $\sim$\crcr\noalign{\kern0.5pt}}}}\limits}

\preprintnumber[4cm]{
UTCCP-P-34\\ April 1998}

\title{
An Introduction to Finite Temperature 
Quantum Chromodynamics on the Lattice%
\thanks{Lectures presented 
at the 1997 Yukawa International Seminar (YKIS'97) on 
``Non-Perturbative QCD --- Structure of the QCD Vacuum ---'',
Kyoto, Japan, 2--12 Dec.\ 1997. To be published in the proceedings
[Prog.\ Theor.\ Phys.\ Suppl.]}
}

\author{
Kazuyuki {\sc Kanaya}
}

\inst{
Center for Computational Physics and Institute of Physics,\\ 
University of Tsukuba, Tsukuba, Ibaraki 305-8577, Japan
}

\recdate{
\today
}

\abst{
In these lectures, we introduce finite temperature QCD on the lattice 
to non-experts of the subject. 
We first formulate lattice QCD both at zero and finite temperatures.
Then a section is devoted to the topic of improved lattice actions
which are becoming an essential ingredient of precision studies
of QCD on the lattice.
We then discuss about finite temperature SU(3) gauge theory, i.e.\
QCD without dynamical quarks (quenched QCD).
Finally, we report recent status of studies in full QCD taking into
account the effects of dynamical quarks.
}
\begin{document}

\maketitle

\section{Introduction}
\label{sec:intro}

Because of the asymptotic freedom in the UV region, 
quantum chromodynamics (QCD) is strongly coupled in the IR region.
Therefore, it is difficult to study the vacuum structure of QCD 
by perturbation theory.
Two characteristic properties of the QCD vacuum are 
quark confinement and the spontaneous breakdown of the chiral symmetry.
On the other hand, IR properties are affected by temperature. 
Actually, studies on the lattice have shown that, when the temperature 
becomes sufficiently high, the low temperature hadronic phase of QCD
turns into the high temperature quark-gluon-plasma phase, in which
quarks are liberated and chiral symmetry is restored. 
The quark-gluon-plasma phase is expected to be realized in the early Universe 
and also possibly in heavy-ion collision experiments.
A non-perturbative study is required to understand the QCD vacuum 
at finite temperatures.

In a conventional formulation of quantum field theories, we first have to
regularize divergences appearing in a perturbative expression of loop 
corrections, and then perform renormalization, order by order in 
perturbation theory, in order to obtain finite results by removing the 
divergences.
This formulation is deeply based on the perturbation theory 
and can not be applied to a non-perturbative study.
Therefore, in order to investigate QCD at finite temperatures, 
we need a definition of QCD that does not resort to perturbation theory.

This leads us to introduce the lattice as a non-perturbative
regularization of the theory:
When we define field variables on a 4-dimensional hyper-cubic 
lattice with the lattice spacing $a$,
the Fourier transforms of lattice fields are periodic in momentum space, 
so that we can restrict all momenta to the first Brillouin zone
$ 
-\pi/a < p_\mu \leq \pi/a.
$ 
Therefore, a finite lattice spacing $a$ naturally provides us with an 
UV cut-off of $O(1/a)$.

When the lattice volume is also finite, the theory is finite and well-defined.
Of course, the continuum field theory is obtained in the double limits of 
infinite lattice volume and zero lattice spacing.
In these limits, the IR and UV divergences are recovered.
Before taking these limits, however, we can introduce different calculation 
techniques.
Therefore, a procedure for a non-perturbative calculation of field theory 
is as follows:
\begin{enumerate}
\item Formulate the model at finite lattice spacing $a$ and finite 
lattice volume $V$.
\item Perform non-perturbative calculations.
\item Take the limits $a\rightarrow 0$ and $V\rightarrow \infty$.
\end{enumerate}
For a non-perturbative calculation in the second step, 
we can perform numerical simulations using super-computers,
as well as analytic studies using strong coupling expansions etc..
Numerous developments and ideas in the last two decades both 
in algorithms for numerical computations and in the computer technology, 
have made lattice field theory one of the most powerful tools to compute 
the non-perturbative properties of QCD.

In these lectures, we attempt to introduce the basic formulation of
lattice QCD and its applications to finite temperature physics.
In Sec.~\ref{sec:formulation}, QCD is formulated on the lattice
both at zero and finite temperatures.
In Sec.~\ref{sec:improvement}, recent developments in improving the 
lattice action are discussed.
Sec.~\ref{sec:puregauge} is devoted to the results in finite temperature
pure gauge theories.
Finally, in Sec.~\ref{sec:fullQCD}, the recent status of finite temperature 
QCD simulations with dynamical quarks is discussed.
A brief summary is given in Sec.~\ref{sec:summary}.

There are a few standard text books on lattice gauge theories\cite{LGT}.
The development in the field is quite fast.
The status of lattice QCD is summarized in reviews in the 
Proceedings of recent international symposium on lattice field theory
``Lattice XX'' \cite{LatticeXX}.

\section{Formulation of QCD on the lattice}
\label{sec:formulation}

\subsection{Euclidian field theory}
Lattice field theory is based on a path-integral representation of 
the field theory.
\be
Z= \int [{\rm d}\phi] \, e^{iS[\phi]},
\ee
where $S[\phi] = \int d^3{\bf x} dt L(\phi,\partial_\mu\phi)$ 
is the action.
In order to apply powerful techniques developed in statistical mechanics, 
we consider the theory in the euclidian space-time 
by substituting $t$ with an imaginary time, $t \rightarrow -ix_4$, 
with real $x_4$.
\be
Z= \int [{\rm d}\phi] \, e^{-S_E [\phi]},
\label{eq:Zeuclid}
\ee
where $S_E = -iS$ is the euclidian action.
In the euclidian space-time, the propagator of a free particle is 
just an exponential function in the coordinate space, 
where the mass appears as inverse correlation length.
In the following sections, we drop the suffix $E$ from the euclidian action.

Lattice discretization of the Euclidian space-time in (\ref{eq:Zeuclid}) 
defines the lattice field theory we shall study.
For definiteness, we consider 4-dimensional hyper-cubic lattices,
unless stated otherwise.
The lattice points are called ``sites'' and the bonds connecting
the nearest neighbor sites are called ``links''.

Matter fields are introduced on the sites. Then,
the simplest lattice action can be obtained by replacing the derivatives 
in the continuum euclidian action by lattice differentiations:
$ 
\partial_\mu \phi(x) \longrightarrow \frac{1}{2a}
\left[\phi(x + \hat\mu) - \phi(x - \hat\mu)\right],
$
where $\hat\mu$ is the lattice unit vector in the $\mu$-th direction
with the length $a$.
In the limit $a \rightarrow 0$, the lattice action smoothly 
recovers the continuum action.
The objective of a lattice field theory is to compute the 
continuum limit of expectation values:
\be
\langle {\cal O} \rangle = {1\over Z} \int [{\rm d}\phi]\, {\cal O}[\phi] \, 
 e^{-S[\phi]},
\hspace{8mm} 
Z = \int [d\phi] \, e^{-S[\phi]}.
\label{eq:expectationValue}
\ee

\subsection{Gauge theory}
In the continuum, a gauge field $A_\mu(x)$ is introduced to intermediate 
the different local gauge transformations at infinitesimally neighboring 
points $x$ and $x+{\rm d}x_\mu$.
Therefore, $\phi^{\dagger} \partial_\mu \phi$, for example, becomes invariant
when we substitute $\partial_\mu$ by the covariant derivative that contains
$A_\mu(x)$.
When the two points are apart by a finite distance, we should consider 
the ``connection'', defined as a path-ordered integration of 
$A_\mu$ along a path connecting these points:
\be
 U(x,y) = P.O. \, \exp \left[ig\int_{{\rm path:}\, x \rightarrow y} 
 {\rm d}z_\mu A_\mu(z)\right].
\label{eq:connection}
\ee
Under a local gauge transformation 
$\phi(x) \rightarrow V(x)\, \phi(x)$, 
with $V(x)$ an element of the gauge group, the connection transforms as 
\be
U(x,y) \; \longrightarrow \; V(x) \, U(x,y) \, V^\dagger (y),
\ee
so that $\phi$'s at two points can form an invariant by
\be
\phi^\dagger(x)\, U(x,y)\, \phi(y).
\ee

Therefore, 
on a lattice where neighboring points are always separated by a finite 
distance, the fundamental variable for gauge degrees of freedom 
is the connection.
The basic formulation of lattice gauge theories was invented by Wilson 
in 1974\cite{Wilson74}.
For sufficiently small lattice spacing, the connection between 
the neighboring sites $x$ and $x+\hat\mu$ is given by
\be
U(x,x+\hat\mu) \equiv U_{x,\mu} = \exp \left[ ig a A_\mu(x) \right].
\label{eq:link}
\ee
We consider $U_{x,\mu}$ to reside on the link connecting 
$x$ and $x+\hat\mu$, and call it the ``link variable''.
The link variables have an orientation. 
A link in opposite direction (the connection from $x+\hat\mu$ to $x$) 
is given by $[U_{x,\mu}]^\dagger$.

In the pure gauge theory, where we have no matter fields acting as 
sources or sinks, 
the link variables must form closed loops in order to be gauge invariant.
This quantity is called the ``Wilson loop''.%
\footnote{
In the SU($N_c$) gauge theory, $N_c$ link variables 
can also join at a site by the totally antisymmetric tensor $\epsilon$.
Closed loops with such joint points are also gauge invariant.
}
The simplest loop is a loop along an elementary square with four links, 
which we call the ``plaquette''.
Therefore, the simplest gauge action consists of plaquettes only.
\be
S_{\rm gauge} = - \beta \sum_{x,\mu\nu} P_{x,\mu\nu},
\label{eq:SgaugePlaq}
\ee
where
\be
P_{x,\mu\nu} = \frac{1}{N_c}{\rm Re}{\rm Tr}\left[
U_{x,\mu}U_{x+\hat\mu,\nu}U_{x+\hat\nu,\mu}^\dagger U_{x,\nu}^\dagger
\right]
\label{eq:plaq}
\ee
is the plaquette at $x$ in the $(\mu,\nu)$ plane.
Using the identification (\ref{eq:link}) and the Hausdorff formula
$ 
e^{\bf x} e^{\bf y} = e^{{\bf x + y + [x,y]}/2 + \cdots},
$ 
we can show that (\ref{eq:SgaugePlaq}) recovers the continuum 
gauge action $\int {\rm d}^4x \frac{1}{N_c}{\rm Tr}\frac{1}{4}F_{\mu\nu}^2$
in the limit of vanishing $a$, 
provided that the coefficient $\beta$ is given by
\be
\beta = 2N_c/g^2.
\label{eq:couplingbeta}
\ee

Because the link variables are elements of the compact gauge group SU($N_c$),
invariant integration (Haar integration) over them is finite.
Therefore, on a lattice, we are not required to introduce gauge fixing.

\subsection{Continuum limit}
\label{subsec:contlim}

By appropriate redefinitions of the fields and the variables 
by means of the lattice spacing $a$, the lattice field theories can be 
defined only by dimensionless quantities. 
Accordingly, on a computer, we have no dimensionful quantities at the beginning.

The scale $a$ is obtained only through a physical interpretation of results
for dimensionless quantities.
A conventional way to fix the scale in QCD is to identify the rho meson 
correlation length $\xi$ with the inverse rho meson mass in the lattice units
$1/m_{\rho} a$. Then the lattice scale is given by
\be
a = { 1 \over 770 \times \xi } \, {\rm MeV}^{-1}.
\label{eq:aMrho}
\ee
Any dimensionful quantity can be used to fix the scale; 
$m_\rho$, the charmonium 1S-1P hyperfine-splitting, etc..
A conventional choice for the case of pure gauge QCD is the string tension 
in the static quark potential.

From the relation (\ref{eq:aMrho}), we see that a smaller $a$ corresponds 
to a larger $\xi$.
In particular, the limit $a \rightarrow 0$ is achieved at 
$\xi \rightarrow \infty$.
We are interested in the modes with correlation length of $O(\xi)$ in this 
limit.

In the coupling parameter space of a lattice theory, 
$\xi \rightarrow \infty$ is achieved at a second order phase transition point.
Physics of IR modes (modes with large $\xi$) near a 2nd order phase 
transition point can be described by the theory of critical phenomena.
In statistical systems, the critical phenomena have the property of 
{\em universality}, 
{\it i.e.}, they do not depend on the details of the microscopic theory.
We, therefore, expect that the continuum limit of a lattice theory is 
independent on the details of the form of the lattice action.
One of many consequences of this universality is the recovery of rotational
symmetry in the continuum limit, because the physics becomes insensitive to
the choice of the lattice orientation.

In QCD, $\xi \rightarrow \infty$ is realized at $\beta \rightarrow \infty$ 
($g \rightarrow 0$) due to the asymptotic freedom.
Because this is the limit of weak coupling, we can compute the scale 
dependence by perturbation theory.
We can formulate the perturbation theory on the lattice similar to the cases 
in the continuum theory, using the identification (\ref{eq:link}).
The result for the scale dependence from lattice perturbation theory 
is given by the beta-function:
\be
a {{\rm d}g\over {\rm d}a} = b_0 g^3 + b_1 g^5 + \cdots
\ee
where the first two coefficients $b_0 = (1/16\pi^2)(11-2N_F/3)$ 
and $b_1 = (1/16\pi^2)^2\,(102-38N_F/3)$ are universal, 
i.e.\ equal to those for the beta-function in the continuum QCD.

Therefore, in asymptotically free theories, the gauge coupling parameter 
dependence of physical quantities near the continuum limit is under control.
This feature is quite important to extract precise predictions from the 
results obtained on lattices with finite $a$.

\subsection{Fermions}
For fermions on the lattice, a complication exists in the formulation.
When we naively discretize the Dirac action in the continuum, we obtain 
the ``naive'' lattice fermion action:
\be
S_{\rm naive} = a^4\,\sum_x \left\{
\sum_\mu \bar\Psi_x \,\gamma_\mu \frac{\Psi_{x+\hat\mu}-\Psi_{x-\hat\mu}}{2a}
+ m_0\, \bar\Psi_x \Psi_x \right\},
\label{eq:NaiveFAction}
\ee
where $\Psi_x$ is a 4-component Grassmann field residing on the site $x$.
Then the propagator shows the pole for $p_4 = -iE$ at
\be
\sinh^2 Ea = m_0^2 a^2 + \sum_{i=1}^3 \sin^2 (p_i a)
\label{eq:naivePole}
\ee
Near the origin of the momentum space, (\ref{eq:naivePole})
predicts the energy eigenvalues to be
$ 
 E = \pm \sqrt{m_0^2 + \sum_i p_i^2}
$ 
as expected.  However, in addition to this mode,
we also have 7 extra low energy modes inside the first Brillouin zone,
at momenta $\vec{p}\approx(\pi/a,0,0)$, $(0,\pi/a,0)$, $\cdots$,
i.e.\ $\pi/a$ for more than one spatial components of $\vec{p}$.
In the continuum limit, these additional modes also contribute to the 
partition function and the expectation values.
In other words, 
they survive as relevant dynamical freedoms in the continuum limit.
These unwanted modes are called ``doublers''.

The problem of doublers is essentially caused by the fact that 
the derivative in the kinetic term of the Dirac action is first order.
This causes the term $\sum_i \sin^2 (p_i a)$ in (\ref{eq:naivePole})
to vanish not only at the origin but also also at 
$\vec{p}=(\pi/a,0,0)$ etc.
Even if we introduce a more complicated lattice derivative than the
naive action (\ref{eq:NaiveFAction}), the
corresponding lattice derivative in the momentum space changes sign 
from negative to positive near the origin of the momentum space.
Because the same happens at the origins of all other Brillouin zones, 
as far as we assume analytic continuity, 
the lattice derivative necessarily crosses zero again within the first 
Brillouin zone.
This leads to the additional low energy modes.

There exists a rigorous statement, the No-Go theorem by Nielsen and 
Ninomiya\cite{NielsenNinomiya}, that as far as we require 
hermiticity, locality and chiral symmetry, doublers are inevitable.
Therefore, in order to formulate a lattice theory for one flavor of Dirac 
fermions in the continuum limit,
we have to abandon some of these nice properties.
The violated properties should be recovered in the continuum limit.
In the continuum limit, we naively expect that different fermion formalisms
lead to a universal continuum limit, recovering all the violated properties.
Off the continuum limit, it is important to check the lattice artifacts
due to the formulation of lattice fermions.

There have been a few proposals to formulate fermions on the lattice.
Two of them are commonly used in major simulations; 
the Wilson fermion formalism\cite{Wilson77} and 
the staggered (Kogut-Susskind) fermion formalism\cite{Susskind77}.
In the Wilson fermion formalism, the chiral symmetry is violated on 
a finite lattice,
and in the staggered fermion formalism, locality is violated when we 
try to define a single flavor fermion.%
\footnote{
Recently, several new ideas to construct a chiral fermion have been studied.
See reviews on the issue for details\cite{LatticeXX}.
}

\subsubsection{Wilson fermion}
In the formulation of the Wilson fermion\cite{Wilson77}, 
we introduce a second-derivative term, the ``Wilson term'', to the action:
\be
a \cdot a^4\,\sum_{x,\,\mu} \bar\Psi_x \,
\frac{\Psi_{x+\hat\mu}+\Psi_{x-\hat\mu}-2\Psi_{x}}{2a^2}
=
a^{-4} \sum_{p,\,\mu} \bar\Psi_p \frac{1-\cos (p_\mu a)}{a} \Psi_p
\label{eq:WilsonTerm}
\ee
that is of order $a$ for the mode at the origin $p \sim 0$, 
but acts as an $O(1/a)$ mass term to the doublers.
Therefore, the doublers are decoupled in the limit $a \rightarrow 0$,
leaving the physical mode at the origin untouched.
Introducing a dimensionless field $\psi_x = a^{3/2} \Psi_x / \sqrt{2K}$,
with $K=1/(8+2m_0a)$, the Wilson fermion action is customarily written as
\bea
S_{\rm Wilson} &=& \sum_{x,y} \bar\psi_x\, D_{x,y}\, \psi_y 
\label{eq:WilsonFermion} \\
D_{x,y} &=& \delta_{x,y} - K \sum_\mu
\left\{ (1-\gamma_\mu)\, \delta_{x+\hat\mu,y} 
 + (1+\gamma_\mu)\, \delta_{x,y+\hat\mu} \right\}
\label{eq:WilsonKernel}
\eea
The parameter $K$ is called the hopping parameter, and corresponds to 
the freedom of the bare fermion mass:
\be
m_0 =\frac{1}{2a} \left( \frac{1}{K} - \frac{1}{K_c} \right),
\ee
with $K_c = 1/8$ the point where the bare mass $m_0$ vanishes.

In a gauge theory, the kernel $D_{x,y}$ is modified as
\be
D_{x,y} = \delta_{x,y} - K \sum_\mu
\left\{ (1-\gamma_\mu)\, U_{x,\mu} \, \delta_{x+\hat\mu,y} 
 + (1+\gamma_\mu)\, U_{y,\mu}^\dagger \, \delta_{x,y+\hat\mu} \right\},
\label{eq:WilsonUKernel}
\ee
where $U_{x,\mu}$ is the link variable.

For $N_F$ flavors of quarks, we simply sum up quarks with different flavors.
Clearly, the flavor symmetry is manifest.
However, because the Wilson term is essentially the mass term for doublers, 
the chiral symmetry is violated even in the limit of vanishing bare quark mass.
As a result, the global flavor/chiral symmetry of continuum QCD,
${\rm SU}(N_F)_{\rm L}\times{\rm SU}(N_F)_{\rm R}\times{\rm U}(1)_{\rm V}$, 
is explicitly broken down to ${\rm SU}(N_F)_{\rm V}\times{\rm U}(1)_{\rm V}$.

Through a perturbative study of axial Ward identities, it is shown that 
the effects of chiral violation due to the Wilson term are removed 
by appropriate renormalizations, 
including an additive renormalization of the quark mass,
i.e.\ by a shift of $K_c$ as a function of $\beta$ \cite{Bochicchio}. 

Away from the perturbative region, we can define $K_c$ as the points
where the pion mass vanishes at zero temperature,
or alternatively where the quark mass vanishes at zero temperature.
Here, the quark mass is defined through an axial-vector Ward 
identity\cite{Bochicchio,ItohNP},
\be
2 m_q \, Z_P \, \langle\,0\,|\,P\,|\,\pi(\vec{p}=0)\,\rangle
= - m_\pi \, Z_A \, \langle\,0\,|\,A_4\,|\,\pi(\vec{p}=0)\,\rangle
\label{eq:mq}
\ee
where $P$ is the pseudoscalar density and $A_4$ the fourth
component of the local axial vector current, with $Z_P$ and $Z_A$ 
renormalization factors.%
\footnote{
We can alternatively define the quark mass by the perturbative 
formula $m_q = (Z_m/2a)\,(K^{-1} - K_c^{-1})$.
Although these different definitions of $m_q$ give different values
at finite $\beta$, it is shown that they converge to 
the same value in the continuum limit.\cite{CPPACSq}
}
Both definitions give the same $K_c$ when
the chiral symmetry is spontaneously broken\cite{ourStandard96}.
$K_c$ forms a smooth and monotonic curve connecting 
$1/8$ in the weak coupling limit ($\beta=\infty$) 
and $1/4$ in the strong coupling limit ($\beta=0$).
$K_c$ can be considered as the points where the chiral
symmetry is effectively recovered.

Aoki proposed an alternative interpretation of the massless pion at $K_c$ 
without resorting to the chiral symmetry in the continuum limit\cite{Aoki}.
In this picture, the massless pions are identified with the Goldstone 
modes associated with a second order phase transition, 
and a rich phase structure was predicted at $K > K_c$.
Although the physical region relevant to the continuum limit of QCD
is below the $K_c$-line, understanding the system at $K > K_c$
is useful in studying the phase structure of Wilson quarks, in particular,
at finite temperatures\cite{AokiUkawaUmemura}.
See also discussions in Refs.~\citen{ourStandard96} and
\citen{KanayaLat95}.

\subsubsection{Staggered (Kogut-Susskind) fermion}

In the formulation of the staggered fermion,
in order to reduce the number of doublers, the Grassmann field 
$\chi$ on the sites have only one component\cite{Susskind77}.
\bea
S_{\rm stag} &=& \sum_{x,y} \bar\chi_x \, Q_{x,y}\, \chi_y 
\label{eq:stag}\\
Q_{x,y} &=& m_0 a\, \delta_{x,y} +
\frac{1}{2} \sum_\mu (-1)^{x_1 + \cdots + x_{\mu-1}}
\left\{ U_{x,\mu}\, \delta_{x+\hat\mu,y} 
      - U_{y,\mu}^\dagger\, \delta_{x,y+\hat\mu} \right\}.
\eea
The conventional four component Dirac spinor is constructed collecting
the $\chi$ fields distributed on a hypercube.
Because a hypercube has $2^4$ sites, we end up with 4 flavors of
degenerate Dirac fermions.

In this formalism, the flavor-chiral symmetry 
${\rm SU}(N_F)_{\rm L}\times{\rm SU}(N_F)_{\rm R}\times{\rm U}(1)$
of massless $N_F$-flavor QCD 
is explicitly broken down to 
${\rm U}(N_F/4)_{\rm L}\times{\rm U}(N_F/4)_{\rm R}$ 
due to a flavor mixing interaction at $a>0$.
However, at least a part of the chiral symmetry is preserved.
Therefore, the location of the massless point $m_0=0$ is protected
against quantum corrections by this symmetry.
This makes a numerical analysis of chiral properties much easier 
than in the case of the Wilson fermion for which $K_c$ must be determined 
numerically.
It is also known that the staggered fermion requires less computer 
resources (memory etc.) than those for the Wilson fermion.

The action (\ref{eq:stag}) can describe quarks only when $N_F$ is 
a multiple of 4. 
A usual trick for the physically interesting cases $N_F=2$ and 3, 
is to modify by hand the power of the fermionic determinant in the 
numerical path-integration. 
\be
Z = \int [{\rm d}U_{x,\mu}] \, \{\det Q[U]\}^{N_F/4} 
\, e^{-S_{gauge}[U]}.
\ee
This necessarily makes the action non-local, 
which sometimes poses conceptually and technically difficult problems.

\subsection{Finite temperature}
\label{sec:FTFormulation}
Finite temperature field theory is defined by the Matsubara formalism 
for finite temperature statistical systems:
We consider static problems in thermal equilibrium at temperature $T$.
Instead of the time coordinate, we introduce a coordinate 
ranging from zero to $1/T$, which formally looks like an euclidian time.
(We set $k_B=1$ in the following.)
Then, the canonical partition function $Z$ can be written as
\be
Z = {\rm Tr}\, e^{-H/T} = \int [{\rm d}\phi]\, e^{-S[\phi]},
\label{eq:FTZ}
\ee
with $H$ the Hamiltonian and
$
S[\phi] = \int_0^{1/T}{\rm d}x_4 \int {\rm d}^3 {\vec x}\,
L(\phi,\partial_\mu \phi).
$
This expression of $Z$ is formally equivalent to the path-integral 
representation of the partition function for an euclidian field theory.
The differences are (i) the range of the euclidian time $x_4$ is 
$[0,1/T]$. 
and (ii), according to the trace operation in (\ref{eq:FTZ}), bosonic and 
fermionic fields obey periodic and anti-periodic boundary conditions 
in the euclidian time direction.

A lattice discretization of (\ref{eq:FTZ}) defines a finite temperature 
lattice field theory.
In order to approximately realize the thermodynamic limit, 
the lattice size in the spatial direction must be sufficiently larger 
than the size $1/T$ in the time direction.

Denoting the dimensionless lattice size in the time direction as $N_t$, 
the temperature is given by
\be
T=1/N_t a.
\label{eq:FTTemp}
\ee
In QCD with small $N_F$, $a$ is a decreasing function of $\beta$.
Therefore, when $N_t$ is fixed as in many numerical simulations, 
larger $\beta$ corresponds to higher temperature.

\subsection{Numerical simulations}
\label{sec:MC}

\begin{figure}[tb]
\centerline{
\epsfxsize=12cm \epsfbox{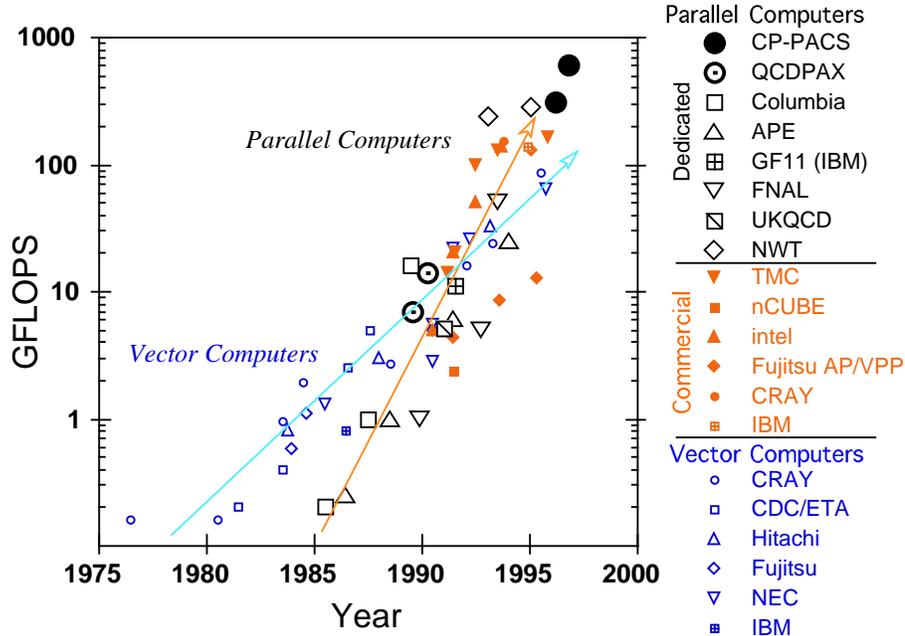}
}
\caption{
Speed of dedicated computers developed by lattice physicists,
together with those of recent commercial machines.
}
\label{fig:Speed}
\end{figure}

When we try to perform a numerical evaluation of the euclidian path integral 
(\ref{eq:expectationValue}), we quickly encounter 
the following two difficulties:
First, the dimension of integration is huge. 
Even on a quite small lattice, for example, $10^4$, 
a 10\,000 dimensional integration is required for each degrees of freedom.
Second, the integrand drastically changes its magnitude, 
forming a quite sharp peak in configuration space. 
This is caused by the factor $e^{-S}$ in (\ref{eq:expectationValue}).
The peaks correspond to the classical solutions and the width of 
the peaks the quantum fluctuations.
Therefore, a naive mesh method to evaluate an integral cannot give 
a reliable value until the mesh becomes extremely fine.
These difficulties can be solved by introducing the Monte Carlo method:
In order to evaluate the integral, we generate the sample points 
with a probability proportional to $e^{-S}$, 
in place of the mesh points in the naive method. 
In the text books, we can find several techniques to generate sample points 
with the correct weight;
the heat-bath method, the Metropolis method, the Langevin method, etc.
The number of sample points required for a given accuracy is much smaller 
than that required in the mesh method, and has no strong dependence on the 
dimension of the integration.

Nevertheless, when we want to obtain a precise result, which is indispensable
in phenomenological studies, the required computer power is enormous.
In order to perform a reliable extrapolation of physical quantities 
to the continuum limit by suppressing lattice artifacts, 
the lattice should be fine enough while keeping a sufficiently big 
physical volume.
Simultaneously, we need a high statistical accuracy which requires a 
large number of sampling points.

In order to generate a configuration (a sample point) in the SU(3) 
pure gauge theory, the conventional pseudo heat-bath 
method\cite{CabibboMarinary}
requires about 5\,700 floating point operations per link.
Vector computers in the '80s and parallel computers in the '90s 
have supported the development of numerical simulations in lattice QCD.
As one of the largest user groups in high performance computing,
lattice physicists have also contributed a lot to the development of 
computer technology itself.
Several groups of lattice physicists have even developed parallel
computers dedicated to lattice QCD\cite{APEmille,QCDSP,cppacs}.
Fig.~\ref{fig:Speed} shows the speed of recent dedicated machines
for lattice QCD.
On the CP-PACS constructed at the University of Tsukuba\cite{cppacs},
more than 10\,000 configurations on a $64^3\times112$ lattice 
can be generated in a day.

When the system includes dynamical fermions, 
different algorithms had to be developed as we cannot have 
Grassmann variables directly on the computers.
However,
even with the latest algorithm\cite{HMC,Ralgo}, several hundred times more 
computer time is required for fermions.
Therefore, major QCD simulations on large lattices have been performed
in the approximation that dynamical pair creation/annihilation of quarks
are neglected (quenched QCD).
Realistic simulations of QCD with dynamical quarks (full QCD) have just 
begun on recent high performance computers, 
by combining big computer power with the idea of improved lattice actions 
which shall be discussed in Sec.~\ref{sec:improvement}.

\section{Improved lattice actions}
\label{sec:improvement}

\subsection{From BETTER to  MUST}
In the scaling region near the continuum limit, we expect that the lattice 
results reproduce the continuum results at distances larger than about the 
correlation length.
However, in general, expectation values at short distances in 
lattice units deviate from the continuum results even at large $\beta$.
Therefore, in order to extract a prediction for the continuum limit,
precise data at large distances on a correspondingly large lattice are
required. Such simulations are quite expensive, in particular, 
near the continuum limit.

\begin{figure}[tb]
\centerline{
\epsfxsize=7cm\epsfbox{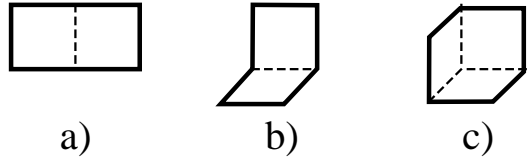}
}
\vspace{-0.2cm}
\caption{Loops of length 6; (a) rectangle, (b) chair,
and (c) parallelogram.
}
\label{fig:loop6}
\end{figure}

On the other hand, due to the universality, 
there are infinitely many candidates for the lattice action
describing the same continuum limit, i.e.\ 
we can introduce additional terms to the standard action 
without affecting the continuum limit.
For example, the lattice gauge action may contain non-minimal loops 
besides the plaquettes adopted in the standard action:
Using loops up to length 6, we can write 
\be
S_{gauge} = - \beta \left\{ c_0 \sum P_{\rm plaq} +c_1 \sum P_{\rm rect}
+ c_2 \sum P_{\rm chair} +c_3 \sum P_{\rm para} \right\}, 
\label{eq:loop6}
\ee
where $P_{\rm rect}$, $P_{\rm chair}$, and $P_{\rm para}$
are rectangle, chair, and parallelogram loops shown in Fig.~\ref{fig:loop6}.
The coefficients in (\ref{eq:loop6}) should satisfy a relation
$c_0 + 8 c_1 + 16 c_2+ 8 c_3 = 1$ in order to obtain the conventional 
continuum gauge action in the limit $a \rightarrow 0$ with the identification 
(\ref{eq:couplingbeta}).
Hence, three parameters are left free.
From the universality, the long distance behavior of the theory is 
insensitive to these additional parameters.
However, short distance properties do depend on these parameters. 
Therefore, a judicious choice of the additional parameters may suppress 
the short distance lattice artifacts even at moderate values 
of the lattice spacing.
Such actions are called ``improved actions''.

Recently, much progress has been reported in improving lattice 
actions\cite{IMreviewH,IMreviewL}.
When an appropriate improved action is obtained,
we will be able to perform a reliable extrapolation 
to the continuum limit using data obtained on coarse lattices.
It is, of course, better to apply such an action to save money 
on computer time.
However, through recent developments in numerical studies of lattice QCD,
we now have stronger motivations to improve the action.

As the first motivation, we would like to introduce the 
recent results of high statistic simulation in quenched QCD 
by the CP-PACS Collaboration\cite{CPPACSq}.
The quenched light hadron spectrum is computed for Wilson quarks 
on lattices with the spatial size $\approx 3$fm 
to an accuracy of 1--3\% in the continuum limit.
The baryon spectrum turns out to show a systematic disagreement 
which is of the order 10\% (maximally 10 standard deviations) 
from the experimental values, which is considered as evidence of 
systematic errors due to the quenching approximation.
Therefore, in order to test QCD to an accuracy better than $O(10\%)$,
we have to perform full QCD simulations without the quenching 
approximation.
Full QCD simulations are, however, extremely computer time consuming 
compared to those of quenched QCD.  
Even with the TFLOPS-class computers that are 
becoming available, high statistics studies, indispensable for reliable 
results, will be difficult for lattice sizes exceeding $32^3\times 64$.
Since a physical lattice size of 
$L\approx 2.5$--3.0fm is needed to avoid finite-size effects,
the smallest lattice spacing one can reasonably reach will be 
$a\sim 0.1$fm.  Hence lattice discretization errors have to 
be controlled with simulations carried out 
at lattice spacings larger than this value.  
This will be a difficult task with the standard plaquette 
and Wilson quark actions 
since discretization errors are of order 20--30\% even at 
$a\approx 0.1$fm. 

We also encounter difficulties with the standard action in a study 
of finite temperature QCD.
In the simulation of a finite temperature system,
the spatial lattice size should be sufficiently large in order
to approximately realize the thermodynamic limit.
With the limitation of the computer power, this means that
we have to suppress the lattice size in the time direction $N_t$.
Currently $N_t \approx 4$--6 is used in most simulations 
with dynamical quarks.
Because the temperature is given by $T=1/N_t a$, 
the corresponding lattices are coarse;
for $T \sim T_c \approx 100$--200MeV, $a \sim 0.2$--0.4fm.
Subsequent lattice artifacts sometimes make the analysis and
interpretation of lattice results not a straightforward task.

The basic idea behind improvement may be obtained by considering the
lattice derivative. The naive derivative
$\Delta_x f(x) = {1 \over {2a}} [f(x+a)-f(x-a)]$
converges to the continuum derivative with errors of $O(a^2)$.
We can reduce this error down to $O(a^4)$ by replacing
$\Delta_x \rightarrow \Delta_x - {a^2 \over 6}\Delta_x^3$
which operates on fields at up to next nearest neighbor sites.
Similar substitution is effective also to obtain a lattice action
that approaches to the continuum action much smoothly.
Here, however, what we want to obtain is not a smoother lattice action, 
but an action which leads to smaller discretization errors in the physical 
observables (\ref{eq:expectationValue}) containing all the 
quantum corrections.

Several different strategies have been proposed to obtain such a lattice 
action.
Two major approaches are the renormalization group (RG) improvement 
programs first proposed by K.G.\ Wilson\cite{Wilson80}, 
and the perturbative improvement programs started 
by K.\ Symanzik\cite{Symanzik83}.
In the following subsections, I describe these methods in more detail. 
Irrespective to the differences in approach, 
the final goal of improvement is to obtain a lattice action
which shows scaling from a coarser lattice.

In general, improved lattice actions contain additional terms
(interactions that usually have a wider spatial size)
compared with the standard action.
Because such additional terms
make the simulation quickly difficult and computer time consuming,
the efficiency of improvement should be tested for each of the new
terms introduced.

\subsection{Symanzik improvement}

In Symanzik's method, we improve physical quantities using 
a result of perturbative computations\cite{Symanzik83,LuescherWeisz85}.
It consists of the following procedures:

\begin{description}
\item[({\it i})] \hspace{2mm} 
Compute a set of physical quantities in perturbation theory 
using a lattice action with non-minimal interactions.
\item[({\it ii})] \hspace{1mm} 
Adjust the non-minimal coupling parameters to remove the 
leading finite $a$ deviations from the continuum limit in these physical 
quantities.
\end{description}
A conventional choice for the physical quantities in this program 
are the low-dimensional operators in the effective action.
When discretization errors are eliminated up to the $n$th order in $a$, 
the resulting action is said to be ``$O({a^n})$ improved''.
For example, the standard plaquette gauge action has $O(a^2)$ errors.
Several different actions removing these errors, to tree 
or one-loop level, have been proposed depending on the choice 
of additional terms\cite{SymanzikAct,SymanzikAct0,SymanzikAct1}.

This improvement program is quite attractive because an improved action 
can be obtained by an analytic calculation.
Nevertheless, the method in its naive form remained unsuccessful for a 
long time. The reason is that the perturbation theory, 
using the bare gauge coupling constant as the expansion parameter,
has poor convergence, and the perturbative results do not agree with
the results from numerical simulations 
which are mostly performed at $\beta=O(1)$.

The failure of the bare perturbation theory may be understood as follows.
The bare perturbation theory is based on an expansion of the link variable
(\ref{eq:link})
\be
U_{x,\mu} \approx 1 + iga A_\mu(x) 
- \frac{1}{2}\,g^2 a^2 \,A_\mu(x)^2 + \,\cdots .
\label{eq:linkexp}
\ee
In order that the perturbation theory works well,
the higher order terms in (\ref{eq:linkexp}) should be much smaller
than 1.
However, in actual simulations, the link variable deviates much from 
unity; we obtain the plaquette expectation value about 0.4--0.6.
Perturbatively, this deviation is caused by large contributions of 
``tadpole'' diagrams\cite{LepageMackenzie}:
By fixing the gauge appropriately, the mean value of the link variable 
can be expressed as
\be
u_0 = \langle U_{x,\mu} \rangle 
\approx 1 - \frac{1}{2}\,g^2 a^2\,\langle A_\mu(x)^2\rangle + \,\cdots.
\label{eq:meanlink}
\ee
Because of a quadratic divergence in the tadpole contribution 
$\langle A_\mu(x)^2 \rangle$, higher order terms 
$[g^2 a^2  \langle A_\mu(x)^2 \rangle]^k$ are suppressed only 
by $g^{2k}$, instead of the naive factor $g^{2k} a^{2k}$.
In the simulations, $g^2 \sim O(1)$.

\subsubsection{Mean-field improvement}

The convergence of the perturbation theory can be improved by an
appropriate choice of the expansion parameter.
Recent significant progress in Symanzik improvement is based 
on a combination of the Symanzik improvement program 
with an idea of ``mean-field improvement (tadpole improvement)''
of the perturbation theory\cite{MFimprove,LepageMackenzie}:
Consider a modified expansion of the link variable around 
the mean-field value $u_0$:
\be
U_{x,\mu} = u_0 \exp \left[ig_{\rm MF}\, a A'_\mu(x) \right].
\label{eq:linkmf}
\ee
With much smaller quantum fluctuations, we will obtain a much better
converging perturbation theory with $A'_\mu(x)$.
From a substitution 
\be
S_{\rm gauge} = - \beta \sum_{\rm plaq}P_{\rm plaq} 
\;\; \longrightarrow \;\; 
- u_0^4 \, \beta \sum_{\rm plaq} \frac{1}{u_0^4}\, P_{\rm plaq},
\ee
we find $g_{\rm MF} = g/u_0^2$.
A conventional choice for $u_0$ is $u_0 = \langle P \rangle^{1/4}$.
The perturbation theory using $g_{\rm MF}$ as the expansion parameter is 
shown to agree with numerical results much more precisely 
even at the large bare coupling $g$ used in numerical 
simulations\cite{LepageMackenzie,IMreviewL}.

Therefore, a prescription for a mean-field improved Symanzik 
improvement is to introduce a third step:

\begin{description}
\item[({\it iii})] \hspace{1mm} 
Multiply an appropriate power of $u_0$ to
the perturbatively determined coefficients of the Symanzik action.
\end{description}

\noindent
It is, however, not trivial whether the non-perturbative quantities we
are interested in are also sufficiently improved by this method.
Non-perturbative tests are required to confirm the efficiency.

\subsubsection{Symanzik-improved Wilson quark action}

The Wilson quark action has $O(a)$ errors.
Symanzik-improved Wilson quark action was studied by Sheikholeslami 
and Wohlert\cite{clover}.
The $O(a)$ improved action reads
\be
S_{\rm clover} = S_{\rm Wilson} 
+ \sum_{x,\mu,\nu} \frac{i}{2}\, c_{\rm SW} K\, \overline{\psi}_x
\sigma_{\mu,\nu} F_{x,\mu\nu} \psi_x,
\label{eq:clover}
\ee
where $\sigma_{\mu,\nu} = \frac{1}{2}[\gamma_\mu,\gamma_\nu]$.
Here $F_{x,\mu \nu}$ is the field strength on lattice.
Because we conventionally construct $F_{x,\mu \nu}$ in terms of 
four plaquette-like loops in the $\mu\nu$ plane around the site  $x$,
the action (\ref{eq:clover}) is called the ``clover action''.

The coefficient $c_{\rm SW}$, the clover coefficient, is 1 at the
tree-level. 
Its mean-field improvement can be done by 
$c_{\rm SW} \rightarrow c_{\rm SW}/u_0^3$ 
because the field strength gives the factor $u_0^{-4}$ while 
the hopping parameter is redefined as 
$K\rightarrow K/u_0$\cite{LepageMackenzie,Mackenzie}.

\subsubsection{Non-perturbative determination of a Symanzik action}

Recently an approach to combine the Symanzik method with a non-perturbative
study was proposed:

\begin{description}
\item[({\it i}$'$)]  \hspace{2mm} 
Measure a set of physical quantities by a numerical 
simulation using a perturbatively inspired form of the Symanzik action, 
varying the coupling parameters freely.
\item[({\it ii}$'$)]  \hspace{1mm}
Adjust the non-minimal coupling parameters of the action
such that several physical requirements are satisfied by the numerical 
data.
\end{description}

\noindent
An attempt to determine $c_{\rm SW}$ non-perturbatively requiring 
a PCAC relation to hold is reported in Ref.~\citen{ALPHA}.
A technical background for this is the development of the Schr\"odinger
functional method in which the lattice correction to the PCAC relation is 
sensitive to the value of $c_{\rm SW}$.

\subsection{RG improvement}

In order to see the basic idea of RG improvement\cite{Wilson80},
let us consider, as a typical example, a block transformation of 
scale factor 2.
A correlation function $G_{\rm block}(r)$ on the blocked
lattice is related to the corresponding correlation function 
$G_{\rm orig}(r)$ on the original lattice by 
$G_{\rm block}(r) \approx G_{\rm orig}(2r)$, 
where the distance $r$ is measured in lattice units. 
Therefore, when we have the continuum behavior at distances 
$r \simm{>} r_0$ on the original lattice, then the continuum behavior 
is realized at $r_0/2$ on the blocked lattice.
This means that a block transformation improves the action.

\begin{figure}[tb]
\centerline{
\epsfxsize=12cm\epsfbox{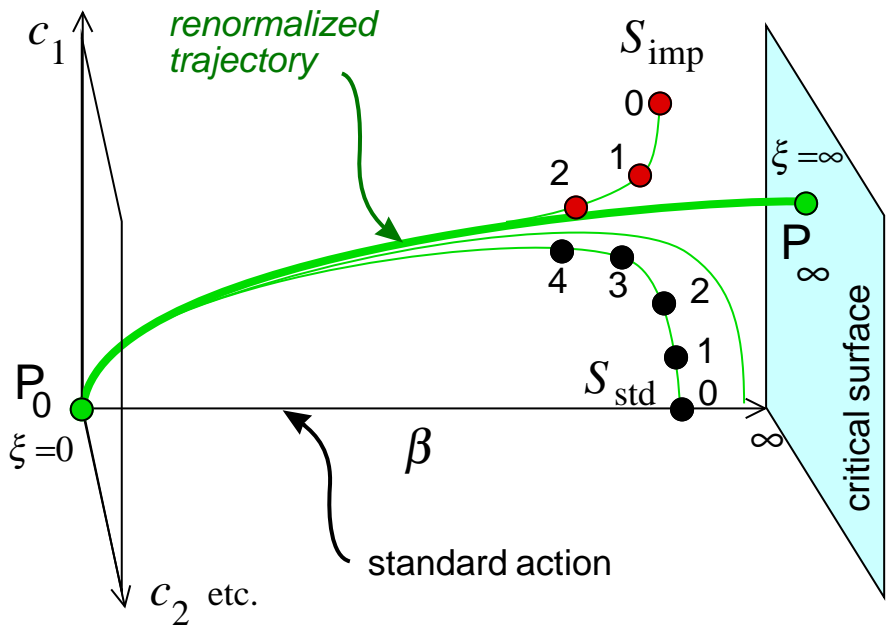}
}
\vspace{-0.1cm}
\caption{RG flow and the renormalized trajectory in the SU(3) 
gauge theory. 
}
\label{fig:RT}
\end{figure}

A block transformation generally induces many effective interactions
in the effective action:
\be
\exp\left\{-S_{\rm block}[\phi_{\rm block}]\right\}
= \int [{\rm d}\phi_{\rm orig}]\, K[\phi_{\rm block},\phi_{\rm orig}]\,
  \exp\left\{-S_{\rm orig}(\phi_{\rm orig})\right\},
\label{eq:blocktransf}
\ee 
where the kernel $K[\phi_{\rm block},\phi_{\rm orig}]$ defines
the block transformation $\phi_{\rm orig} \rightarrow \phi_{\rm block}$.
Therefore, repeated applications of a block transformation defines
a flow (RG flow) in the infinite dimensional coupling parameter space of 
effective actions.

For the SU(3) gauge theory, we imagine the RG flows to be as shown in 
Fig.~\ref{fig:RT}.
In this figure, $c_1$, $c_2$, etc.\ denote additional coupling parameters 
[cf.\ the action (\ref{eq:loop6})].
The points on the $\beta$ axis correspond to the standard one plaquette action
at various $\beta$.
Because a block transformation makes the correlation $\xi$ in lattice units 
smaller, RG flows are oriented towards smaller $\beta$. 
The hyperplane at $\beta=\infty$ ($\xi=\infty$) is called the critical surface.
When the starting action $S^{(0)}$ is in the scaling region that is close to 
the critical surface, 
we expect that all trajectory flows first along the critical surface, and then 
leaves the critical surface to gradually 
approach to a universal curve, the ``renormalized trajectory'' (RT), after 
sufficiently many steps, because the long distance physics is universal in the
scaling region.
The number of block transformations required to get sufficiently
close to RT is nothing but $\log_2$ of the minimum distance
to obtain continuum behavior with $S^{(0)}$.

We expect that the RT is also a RG flow starting from an infra-red fixed 
point, $P_{\infty}$, on the critical surface.
Because the RT is directly connected to the action $P_{\infty}$ in the 
continuum limit by block transformations, actions on the RT show 
continuum properties from the shortest distances on the lattice.
Therefore, the actions on the RT are called 
``perfect actions''\cite{Hasenfratz94}.

If an infinite number of coupling parameters are admitted, a perfect action
is a goal of improvement.
In reality, we can not handle infinitely many couplings.
A few approaches are proposed to obtain an improved action
in a finite dimensional subspace of the coupling parameters.

\subsubsection{Fixed point action approach}

Hasenfratz and Niedermayer proposed to use an approximation of the fixed
point action $P_{\infty}$ as an approximate perfect action\cite{Hasenfratz94}.
In asymptotically free theories, because the critical surface is located 
in the weak coupling region, a compact integral equation for $P_{\infty}$ 
can be written.
With a proper Ansatz for the effective action, we can numerically solve the
fixed point action $P_{\infty}$.
The most delicate point is the choice of the finite-dimensional Ansatz action
(``parametrization'').
It is noted that generally, a large number of coupling parameters are 
required to achieve a good approximation for $P_{\infty}$.

Many developments are reported in the literature;
see Refs.~\citen{IMreviewH,HasenfratzYKIS} for more details.
For a trial to compute the RT nonperturbatively, see Ref.~\citen{QCDTARO_FP}.
The fixed point action for full QCD is discussed in 
Refs.~\citen{IMreviewH,Wiese97,DeGrand97}. 

\subsubsection{Iwasaki's method}

In an alternative approach, proposed in 1983 by Iwasaki\cite{Iwasaki83},
the practical constraint for the number of coupling parameters 
in numerical simulations is taken more seriously.
In this approach, instead of trying to obtain an approximate perfect action
directly, one attempts to accelerate the approach to the RT 
by taking advantage of the extended parameter space:

\begin{description}
\item[({\it i})]  \hspace{2mm} 
We restrict ourselves to a small dimensional coupling parameter 
space consisting only of interactions which can be 
easily implemented in numerical simulations.
\item[({\it ii})]  \hspace{1mm} 
Find a set of coupling parameters that minimizes the distance to
the RT after {\em a few} block transformations.
\end{description}

\noindent
Because a block transformation induces many effective couplings, the number 
of coupling parameters for the initial action can be quite small.

As an illustration, let us consider an action $S_{imp}$ in the 
two-dimwnsional coupling parameter space $(\beta,c_1)$ 
shown in Fig.~\ref{fig:RT}, and suppose 
that the additional parameter $c_1$ is adjusted such that the RG flow 
gets sufficiently close to RT after, say, 2 block transformations.
This means that, when we simulate the system using $S_{imp}$,
a separation of $2^2a$ is enough to get the continuum properties.

Iwasaki applied this program to the SU(3) gauge theory\cite{Iwasaki83}.
Asymptotic freedom again helps us to compute such $S_{imp}$:
\be
S_{imp} = - \beta \left\{ 
c_0 \sum P_{\rm plaq} +c_1 \sum P_{\rm rect} \right\},
\label{eq:Iwasaki}
\ee
where $c_0 = 1 - 8 c_1$ and $c_1 = - 0.331$ ($-0.293$) to minimize 
the distance to RT after one (two) block transformations.
This action is remarkably simple.  It is easy to write a vectorized
and/or parallelized program for this action.
Good efficiency in removing several lattice artifacts is reported
by the Tsukuba group\cite{ourPot97,CPPACSfull}.
An application to finite temperature QCD with dynamical quarks\cite{ourPRL97}
will be discussed in Sec.~\ref{sec:TsukubaNf2}.

\section{$SU(3)$ gauge theory at finite temperature}
\label{sec:puregauge}

As the first step towards finite temperature QCD, in this section,
let us study the case of the SU(3) pure gauge theory, or, equivalently, 
QCD in the approximation that dynamical creation/annihilation
of quarks are neglected (quenched QCD).
Although quenched QCD is not quite realistic, we can obtain
high statistics data on large lattices in this case.
Therefore, it provides us with a good lesson for full QCD studies.

\subsection{Deconfining transition in SU(3) gauge theory}
In a study of the deconfining transition in pure gauge theories,
it is useful to consider the ``Polyakov loop''\cite{Polyakov}
\be
   \Omega_{{\vec x}} = \frac{1}{N_c}\,
            {\rm Tr} \left(\prod_{t=1}^{N_t} U_{{\vec x},t;4}\right).
\label{eq:Polyakov}
\ee
In the path integral, $\Omega_{{\vec x}}$ is just the factor
from the current appearing when a static charge is located at 
the spatial point ${\vec x}$.
Therefore, except for the renormalization corrections,
$\langle \Omega \rangle \sim e^{-F_q/T}$, where $F_q$ is the
free energy of the static charge.
We expect that $F_q=\infty$ (i.e.\ $\langle\Omega\rangle=0$) 
when charge is confined,
while $F_q<\infty$ ($\langle\Omega\rangle>0$) when charge is deconfined.
Therefore, $\langle\Omega\rangle$ is an order parameter for the deconfining 
transition.

The global symmetry behind this order parameter is given by
\bea
U_{x,\mu} \rightarrow \left\{ 
\begin{array}{ll} z\,U_{x,\mu} & \mbox{if $x_4=0$ and $\mu=4$} \\
                  U_{x,\mu}    & \mbox{otherwise}
\end{array} \right.
\label{eq:center}
\eea
where $z$ is an element of the center group Z($N_c$) of the gauge
group SU($N_c$). 
Because $z$ commutates with any element of SU($N_c$), the plaquettes
as well as more extended closed loops are invariant under the 
transformation (\ref{eq:center}),
i.e.\ the pure gauge actions (\ref{eq:SgaugePlaq}), (\ref{eq:loop6}), etc.\
are invariant under (\ref{eq:center}).
On the other hand, because $\Omega_{{\vec x}}$ crosses the hyper plane
$x_4=0$ only once, $\Omega_{{\vec x}} \rightarrow z\,\Omega_{{\vec x}}$
under (\ref{eq:center}).
Therefore, a non-vanishing vacuum expectation value of $\Omega$ 
implies the spontaneous breakdown of the center $Z(3)$ global symmetry 
at the deconfining transition.

\begin{figure}[tb]
\centerline{
\epsfxsize=14cm\epsfbox{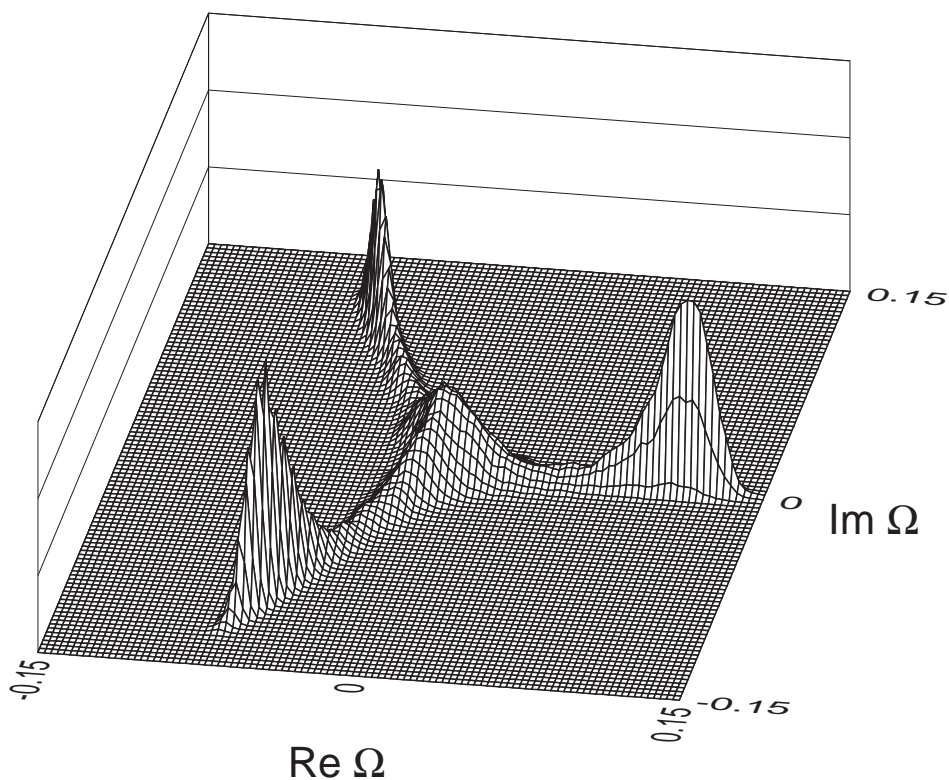}
}
\vspace{-0.7cm}
\caption{
Polyakov loop histogram in the SU(3) gauge theory at the deconfining
transition point obtained on
a $24^2\times36\times4$ lattice\protect\cite{QCDPAX}.
}
\label{fig:PolyHstg}
\end{figure}

\begin{figure}[tb]
\centerline{
a)\epsfxsize=6.2cm\epsfbox{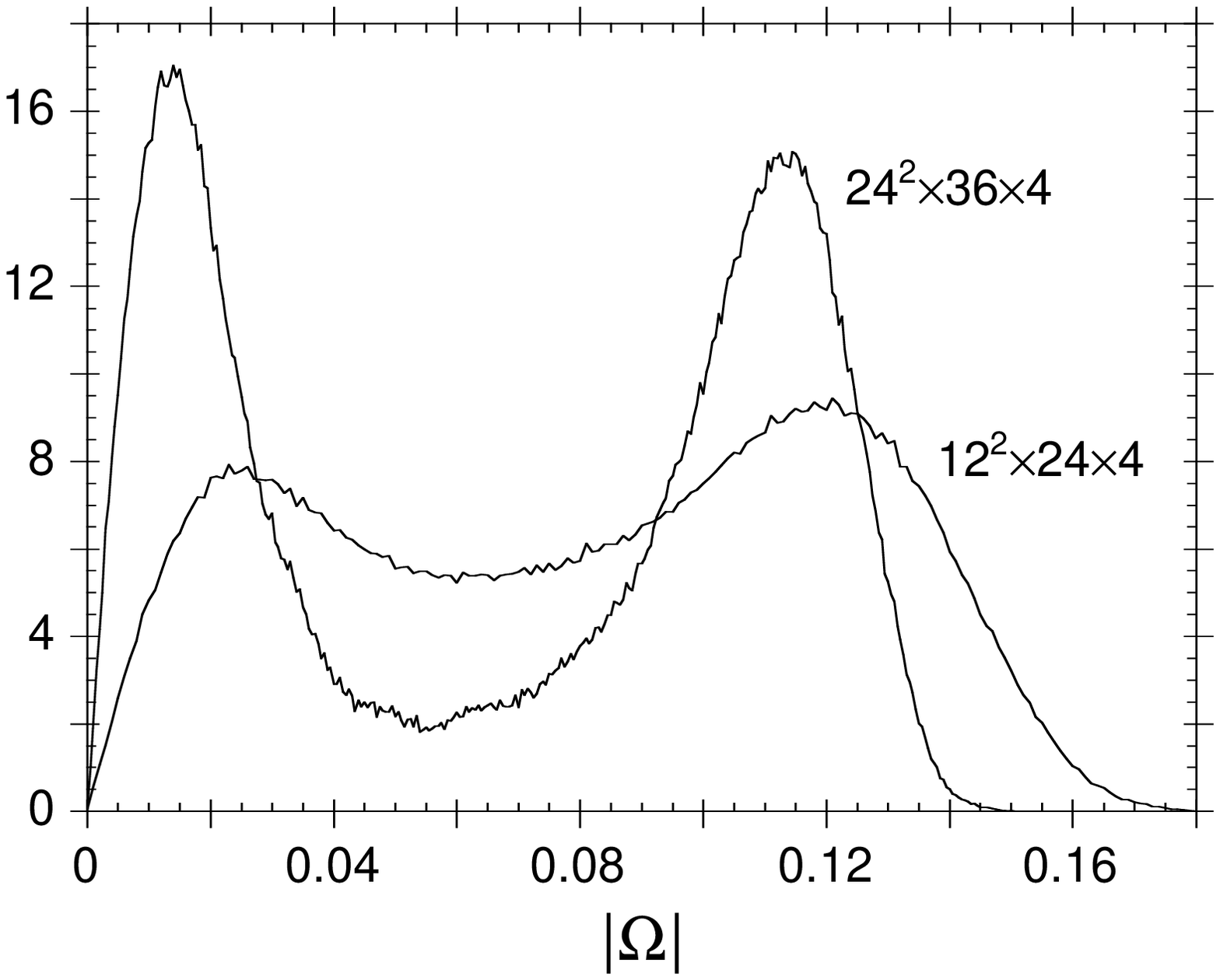}
\makebox[2mm]{}
b)\epsfxsize=6.2cm\epsfbox{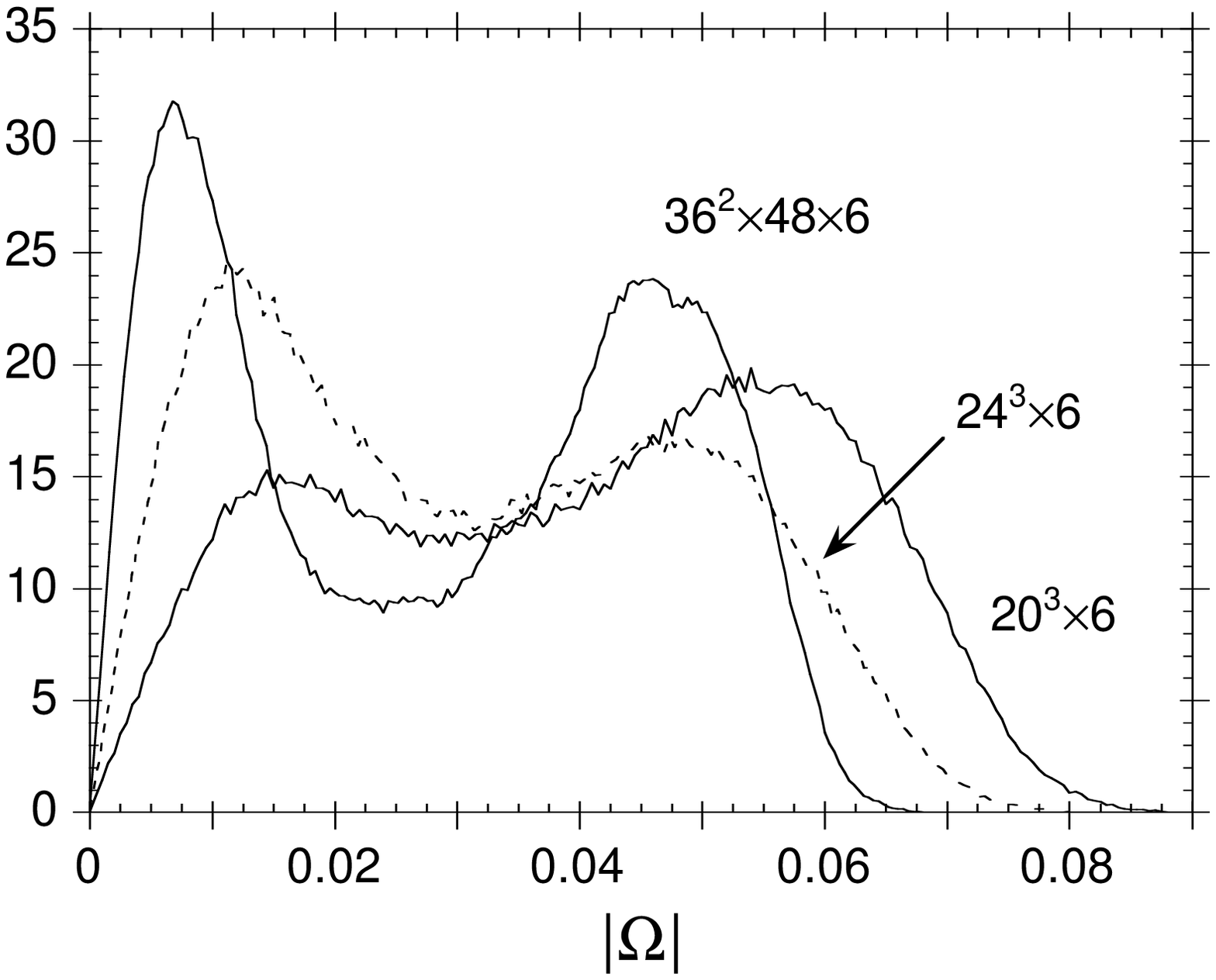}
}
\caption{
Polyakov loop histogram for $|\Omega|$ at the deconfining
transition point obtained on
(a) $N_t=4$ and (b) 6 lattices \protect\cite{QCDPAX}.
The simulation temperature for the $24^3\times6$ lattice is slightly
lower than the transition temperature.
}
\label{fig:PolyAHstg}
\end{figure}

We may study the nature of the deconfining transition using an effective
theory of the Polyakov loops in three dimensions.
Because the order-disorder phase transition in a three dimensional Z(3) 
spin model (the Potts model) is first order, we may expect that the
deconfining transition in the SU(3) gauge theory is also first order,
when the interaction in the effective theory of $\Omega_{\vec x}$
is short ranged\cite{SvetiskyYaffe}.

Figure~\ref{fig:PolyHstg} is a result of the Polyakov loop histogram
in the complex plane, obtained just at the deconfining transition 
temperature in the SU(3) gauge theory\cite{QCDPAX}.
The peak at the center $\Omega \sim 0$ is the contribution of the 
symmetric phase (low temperature confining phase) and the three
peaks at non-zero $|\Omega|$ with $\arg\Omega \sim 0$ and $\pm 2\pi/3$ 
are from the Z(3) broken phases.
Clear separation of these peaks, as well as the lattice volume
dependence shown in Fig.~\ref{fig:PolyAHstg}, implies that 
the transition is of first order. 
The order of the transition is confirmed by a precise finite size
scaling test\cite{FOU90,QCDPAX}.
Accordingly, the short range nature of the effective interaction
between the Polyakov loops is also shown\cite{FOU90}.

\subsection{Transition temperature}

Precisely speaking, we have no real transition on a finite lattice.
The transition point $\beta_c(N_t)$ for each fixed $N_t$ can be
computed by an extrapolation to the infinite spatial volume using 
a finite size scaling.
The value of $\beta_c(N_t)$ is translated into physical units
by fixing the scale at this value of $\beta$.

In pure gauge theories, the scale is conventionally fixed with
the string tension in a static quark potential $\sigma$ 
at zero temperature.
\be
\sigma a^2 = -\lim_{A\rightarrow\infty}\frac{1}{A}\,\ln\langle W(A)\rangle
\ee
where $W(A) = \prod_{\ell\in \partial A} U_\ell$ is the Wilson loop
with area $A$.
We then obtain 
$T_c/\sqrt{\sigma} = 1/[N_t\sqrt{\sigma a^2 (\beta_c(N_t))}]$.%
\footnote{
Ironically, the largest error comes from $\sigma a^2$ determined at 
zero temperature, because an extraction of $\sigma a^2$ contains
many delicate fittings.
The systematic errors from the choice of the fitting ansatz, 
fitting range, etc.\ are not fully estimated.
}
Using recent data\cite{SCRI97} for $\sigma a^2$, we obtain
$T_c/\sqrt{\sigma} \approx 0.63(1)$ in the continuum limit 
($N_t \rightarrow\infty$) from the standard action\cite{QCDPAX,Boyd95}.
Results from various improved actions are 
$T_c/\sqrt{\sigma} \approx 0.62$--0.66 
\cite{ourPot97,DeGrand95,Bliss96,Beinlich97s}.
Adopting a phenomenological value $\sigma=(427\mbox{MeV})^2$
from a charmoniun spectrum\cite{Eichten80},
we obtain $T_c \sim 270$MeV.

\subsection{Thermodynamic quantities}
\label{sec:ep}

In a phenomenological study of the quark-gluon plasma in 
heavy ion collisions and in the early Universe, 
it is important to evaluate thermodynamic quantities such as 
the energy density $\epsilon$ and the pressure $p$, 
near the transition temperature of the deconfining phase transition. 
These quantities are defined by derivatives of the partition function 
with respect to the temperature $T$ 
and the physical volume $V$ of the system
\begin{eqnarray}
\epsilon = - \frac{1}{V}\, \frac{\partial \ln Z}{\partial T^{-1}}, 
\hspace{8mm} 
 p       = T\, \frac{\partial \ln Z}{\partial V}. 
\label{eqn:ep}
\end{eqnarray}
On a lattice with the size $N_s^3\times N_t$, $V$ and $T$ are given by 
$V = (N_s a_s)^3$ and $T = 1 / (N_t a_t)$, 
with $a_s$ and $a_t$ the lattice spacings in spatial and 
temporal directions.
Because $N_s$ and $N_t$ are discrete parameters, 
the partial differentiations in (\ref{eqn:ep}) are performed 
by varying $a_s$ and $a_t$ independently.

Anisotropy on the lattice is introduced by 
different coupling parameters in temporal and spatial directions.
For an SU($N_c$) gauge theory, the standard plaquette action 
on an anisotropic lattice is given by
\begin{eqnarray}
 S = - \beta_s \sum_{x,\, i<j \ne 4} P_{x,ij}
     - \beta_t \sum_{x,\, i \ne 4}   P_{x,i4},
\end{eqnarray}
where $P_{x,\mu\nu}$ is defined by (\ref{eq:plaq}).
Then the energy density and pressure, renormalized at $T=0$, 
are given by\cite{Engels82,Karsch82}
\begin{eqnarray}
\epsilon a_s^4 
&=& 3\xi^2 
  \left\{ \frac{\partial \beta_s}{\partial \xi} \,
 \left( \langle P_s \rangle - \langle P \rangle_0 \right) 
 + \frac{\partial \beta_t}{\partial \xi} \,
 \left( \langle P_t \rangle - \langle P \rangle_0 \right)
 \right\} \label{enrg} \\
p a_s^4 
&=& \xi^2 
  \left\{
  \left( \frac{\partial \beta_s}{\partial \xi}
    +\frac{a_s}{\xi}\frac{\partial\beta_s}{\partial a_s} \right) \,
  \left( \langle P_s \rangle - \langle P \rangle_0 \right) \right.
\nonumber \\ & &
  \left.
 +\left( \frac{\partial \beta_t}{\partial \xi}
    +\frac{a_s}{\xi}\frac{\partial\beta_t}{\partial a_s} \right) \,
  \left( \langle P_t \rangle - \langle P \rangle_0 \right)
 \right\}, \label{prs}
\end{eqnarray}
where $\langle P_{s(t)} \rangle$ is
the space(time)-like plaquette expectation value 
and $\langle P \rangle_0$ the plaquette expectation value 
at the same coupling parameters on a zero-temperature lattice.
In these expressions, the variables $\xi \equiv a_s/a_t$ and $a_s$ are 
chosen to control the lattice spacings.

Therefore, in order to compute $\epsilon$ and $p$ from the results
of simulations, the values for the derivatives of gauge coupling constants
with respect to the anisotropic lattice spacings 
(the anisotropy coefficients) are required.

The calculation of these anisotropy coefficients in the lowest order 
perturbation theory was done by Karsch\cite{Karsch82}.
[Coefficients needed for the case with dynamical quarks are also computed 
perturbatively in Ref.~\citen{Trinchero83}.]
However, the bare perturbation theory is not reliable for the values of 
$\beta$ where MC simulations are performed.
Accordingly, the perturbative coefficients are known to lead to
a pathological result of negative pressure at strong couplings 
used in the numerical simulations.
In the case of SU(3) gauge theory, the transition is of first order.
At a first order transition point, we have a finite gap for the energy 
density, the latent heat, but expect no gap for the pressure.
It is known that the perturbative anisotropy coefficients have the
problem of non-vanishing pressure gap at the deconfining transition
point: $\Delta p / T^4 = -0.32(3)$ and $-0.14(2)$ on
$24^2\times36\times4$ and $36^2\times48\times6$ lattices\cite{QCDPAX}.
Therefore, we need non-perturbative values for the anisotropy 
coefficients.

We are mainly interested in the values of the anisotropy coefficients 
for the case of isotropic lattices 
($\beta_s=\beta_t\equiv\beta$, i.e.\ $\xi=1$)
because most simulations are performed in this case.
At $\xi=1$, we have 
$
\left(a_s \partial \beta_s/\partial a_s\right)_{\xi = 1}
= \left(a_s \partial \beta_t/\partial a_s\right)_{\xi = 1}
= a{\rm d}\beta/{\rm d}a
$,
where $a{\rm d}\beta/{\rm d}a$ is the beta-function.
Furthermore, a combination of the remaining two anisotropy coefficients 
is known to be related to the beta-function\cite{Karsch82}:
$
\left(\partial \beta_s/\partial \xi\right)_{\xi = 1}
+ \left(\partial \beta_t/\partial \xi\right)_{\xi = 1}
= -(1/2)\, a{\rm d}\beta/{\rm d}a.
$
Therefore, 
we need to estimate nonperturbative values of the beta-function 
$a{\rm d}\beta/{\rm d}a$ and 
one independent combination of the anisotropy coefficients.

In connection to this, it is useful to consider a combination $\epsilon-3p$
which is defined via a uniform scale transformation:
\bea
\epsilon - 3p 
&=& - \frac{a^4}{V/T} 
    \left( \frac{1}{T} \frac{\partial}{\partial(1/T)}
    + 3V\frac{\partial}{\partial V} \right) \ln Z 
 =  - \frac{a^4}{V/T} 
    \left( a_t \frac{\partial}{\partial a_t}
    + a_s \frac{\partial}{\partial a_s} \right) \ln Z   \nonumber \\
&=& -3 T^4 N_t^4 \, a\frac{{\rm d}\beta}{{\rm d}a} 
\left(\langle P_s + P_t \rangle - 2 \langle P \rangle_0 \right)
\label{eq:e3p}
\eea
at $\xi=1$.
Therefore, this combination depends only on the beta-function
$a {\rm d}\beta/{\rm d}a$.
Several non-perturbative values for the beta-function are 
available; from a MC renormalization group study\cite{QCDTARO93},
from a study of the transition temperature\cite{Boyd95},
or from the $\beta$-dependence of a physical quantity,
such as the string tension\cite{SCRI97}.

\begin{figure}[tb]
\centerline{
a)\epsfxsize=13cm\epsfbox{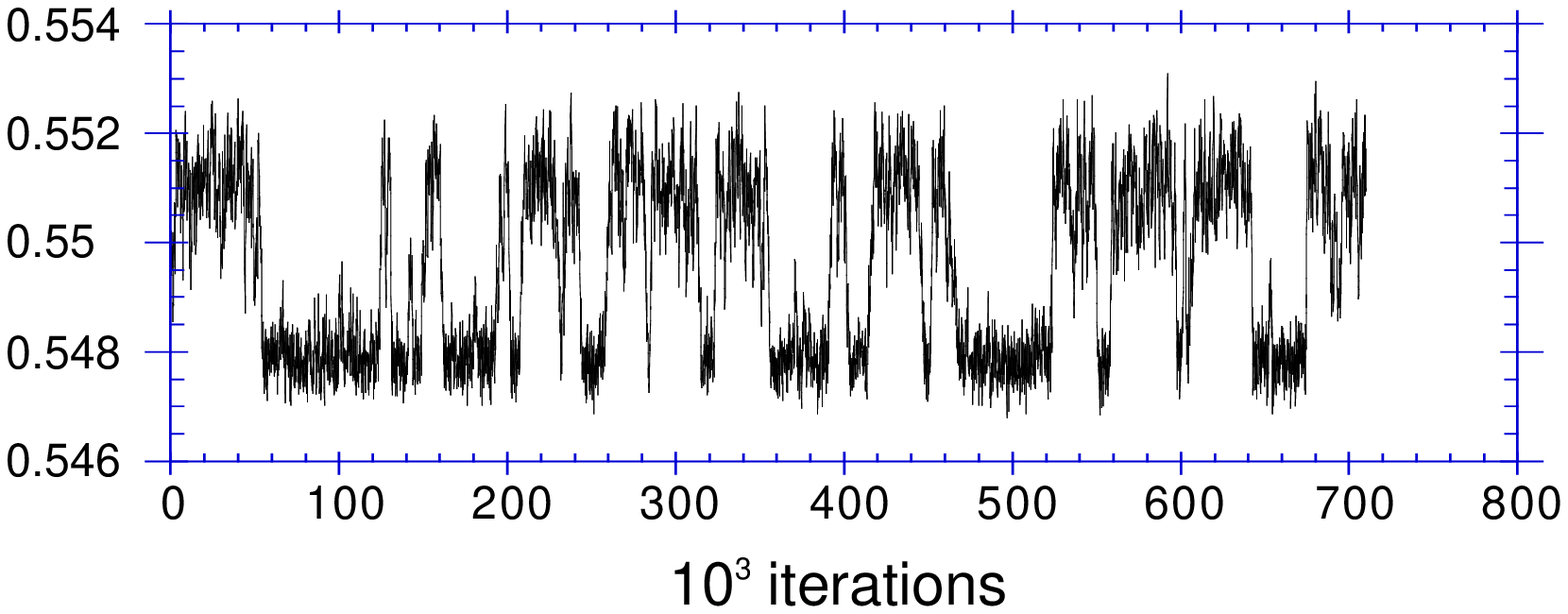}
}
\vspace{5mm}
\centerline{
b)\epsfxsize=13cm\epsfbox{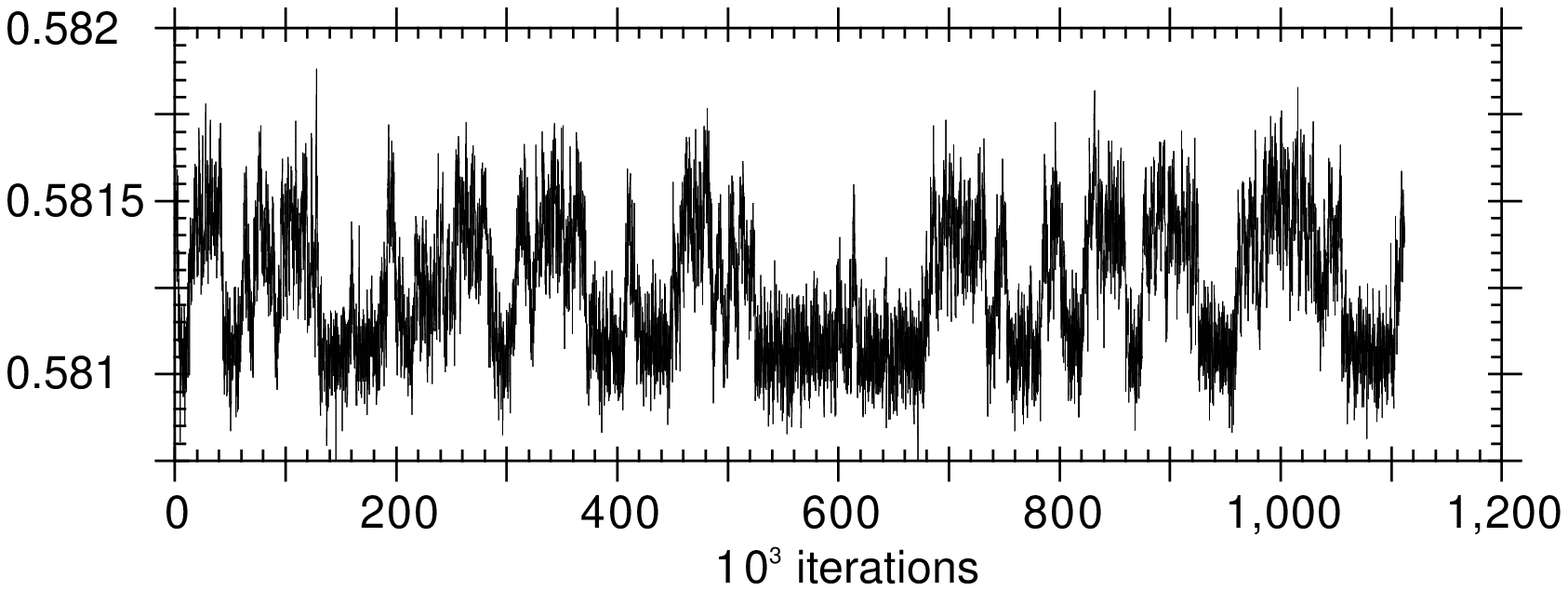}
}
\caption{
Plaquette history in the SU(3) gauge theory at the deconfining
transition point obtained on 
(a) $24^2\times36\times4$ and
(b) $36^2\times48\times6$ lattices\protect\cite{QCDPAX}.
}
\label{fig:PlaqHist}
\end{figure}

\begin{figure}[tb]
\centerline{
\epsfxsize=12cm\epsfbox{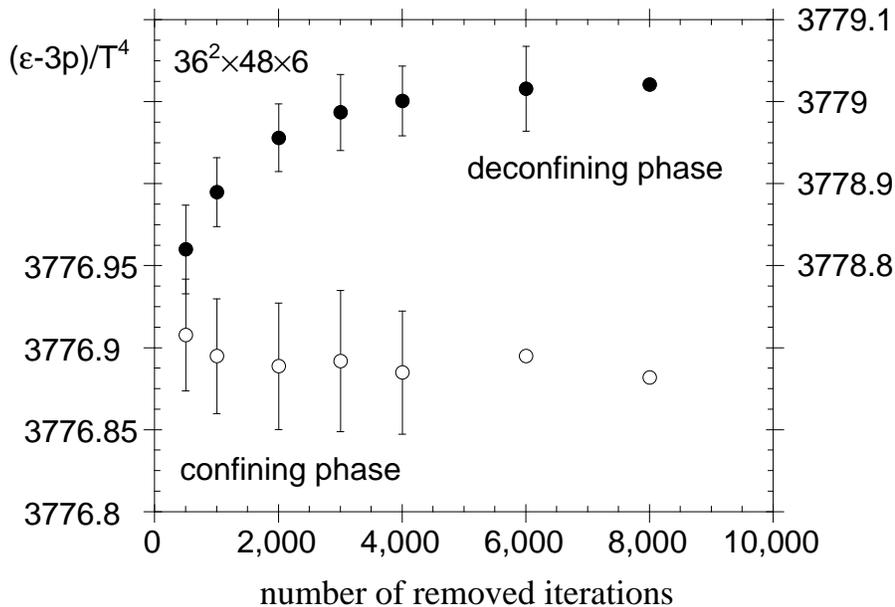}
}
\caption{
$(\epsilon-3p)/T^4$ average in each of the confining and deconfining 
phases in the SU(3) gauge theory at the deconfining transition point 
obtained on a $36^2\times48\times6$ lattice\protect\cite{QCDPAX},
as a function of the removed iterations around the phase flip-flops
shown in Fig.~\protect\ref{fig:PlaqHist}(b).
$\epsilon-3p$ is defined by (\protect\ref{eq:e3p}) using the perturbative 
beta-function.
}
\label{fig:e3p}
\end{figure}

As an application, we compute the energy gap $\Delta\epsilon$,
identified with $\Delta(\epsilon-3p)$ because we expect $\Delta p=0$,
at the deconfining transition in the SU(3) gauge theory.
In order to determine the expectation values in each phase, 
we first inspect the ``flip-flops'' 
between the confining and deconfining phases 
in Monte Carlo time histories (see Fig.~\ref{fig:PlaqHist} for example),
and separate the runs into the two phases by the flip-flops.
A sufficient number of iterations around the flip-flops and around the spikes 
should be removed to avoid contamination from transition stages.
Fig.~\ref{fig:e3p} shows an example of expectation values of observables 
in each phase as a function of the number of removed iterations. 
We can obtain stable results when a sufficiently large number of 
iterations from the transition stages are removed.%
\footnote{
When we try to define an expectation value for a phase in terms of a cut 
for the spatial average of the Polyakov loop etc., 
the result strongly depends on the value of the cut. 
Therefore, we cannot obtain a reliable result in this way.
}
Note that such unambiguous separation of two phases is possible
only on large lattices where the persistence time of each phase
is sufficiently long.
Using the beta-function calculated from recent string tension 
data\cite{SCRI97},
we find that $\Delta(\epsilon-3p)/T_c^4 = 2.072(43)$ and 1.578(42)
for $N_t=4$ and 6, respectively, with the standard action\cite{QCDPAX}.
Using Symanzik improved actions with $2\times1$ loops, 
values of 1.57(12) and 1.40(9) are reported for tree-level and 
tadpole-improved actions on a $32^3\times4$ lattice\cite{Beinlich97}.
More data at larger $N_t$ are needed to make a reliable continuum 
extrapolation.

In order to determine $\epsilon$ and $p$ separately, 
we need one more input as discussed above.
A non-perturbative determination of a combination 
of the anisotropy coefficients was attempted in 
Refs.~\citen{Burgers88,Fujisaki97,Scheideler98,Klassen98}
using a matching of space-like and time-like Wilson 
loops on anisotropic lattices (the matching method)\cite{Burgers88}.
Alternatively, we can evaluate a non-perturbative value of 
pressure directly from the Monte Carlo data 
by the integral method\cite{Fingberg90}:
assuming homogeneity of the system, expected for
the case of large spatial volume, we obtain the relation 
$ p = -f$, where $f$ is the free energy density, 
$ f = - \frac{T}{V} \ln Z$.
For the case of pure gauge theory with the plaquette action, 
$f$ can be evaluated by integrating the plaquette 
$\langle P \rangle$ in term of $\beta$ on isotropic lattices, 
since 
$ \frac{\partial}{\partial \beta} \ln Z = 6 N_s^{3} N_t 
\langle P \rangle$.
The resulting value of the pressure, in turn, provides us with 
a non-perturbative estimate of a combination of the anisotropy 
coefficients\cite{Boyd95,Fingberg90,Beinlich96,Blum95}.
Recently, a new method, using a measurement of the finite temperature 
deconfining transition curve in the lattice coupling parameter space 
extended to anisotropic lattices, was proposed\cite{Ejiri98}.

\begin{figure}[tb]
\centerline{
\epsfxsize=12cm\epsfbox{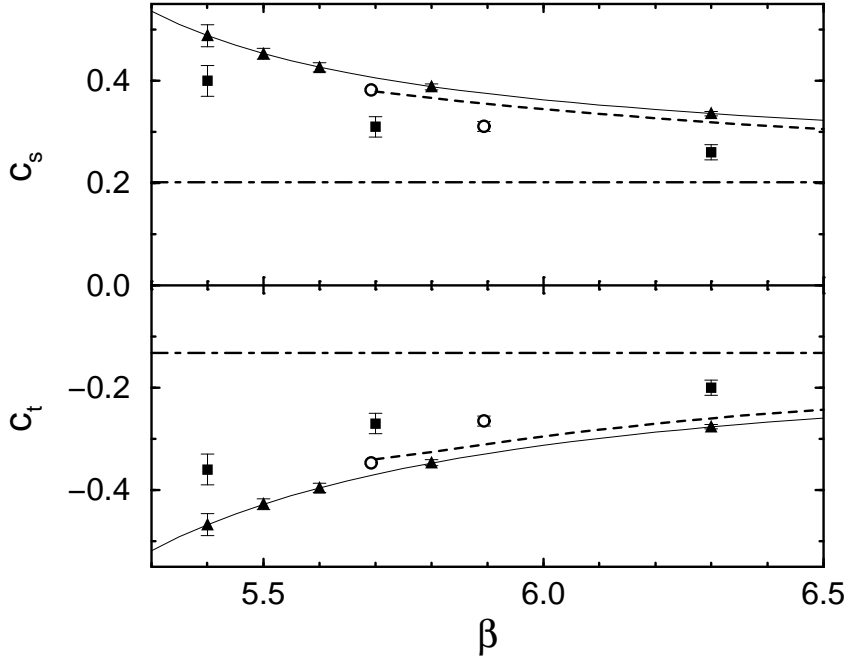}
}
\vspace{-0.4cm}
\caption{
Anisotropy coefficients (Karsch coefficients) in the SU(3) gauge 
theory.
Dot-dashed lines are the results of the perturbation theory 
\protect\cite{Karsch82}.
Filled squares are the results of a matching method obtained in 
Ref.~\protect\citen{Scheideler98}.
Filled triangles and thin lines are the results of a matching method 
\protect\cite{Klassen98} combined with the beta-function 
using a recent string tension data\protect\cite{SCRI97}.
Dashed lines are the results of the integral method \protect\cite{Boyd95}.
Open circles are the results of a new method using the transition curve 
in the coupling parameter space for anisotropic lattices 
\protect\cite{Ejiri98}.
}
\label{fig:CsCt}
\end{figure}

Recent results for the anisotropy coefficients in the form 
\bea
c_s 
= \left(\frac{\partial g_s^{-2} }
    {\partial \xi}\right)_{a_s: {\rm fixed},\, \xi = 1}, 
\hspace{8mm} 
c_t 
= \left(\frac{\partial g_t^{-2} }
    {\partial \xi}\right)_{a_s: {\rm fixed},\, \xi = 1},
\label{eq:CsCt}
\eea
(Karsch coefficients) with
$\beta_s = 2N_{c} g_s^{-2} \xi^{-1}$ and 
$\beta_t = 2N_{c} g_t^{-2} \xi$, 
are summarized in Fig.~\ref{fig:CsCt}.
Applying these results, we can reanalyze the pressure gap.
At the deconfining transition point for $N_t=6$,
we find $\Delta p/T^4 = -0.003(17)$ using the results of 
Ref.~\citen{Ejiri98}
or $-0.040(43)$ using the result of Ref.~\citen{Klassen98}.
We find that the problem of a non-vanishing pressure gap is removed
with non-perturbative anisotropy coefficients.

\subsection{Interface tension}

\begin{figure}[tb]
\centerline{
a)\epsfxsize=7.8cm\epsfbox{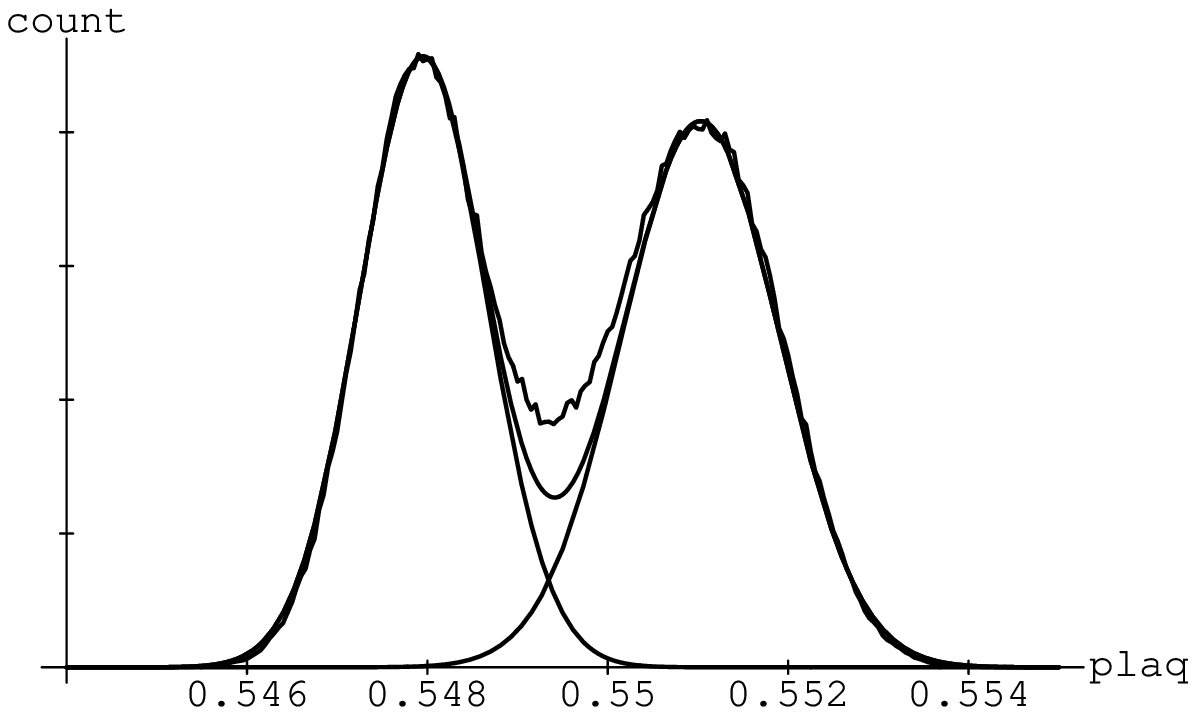}
b)\epsfxsize=5.7cm\epsfbox{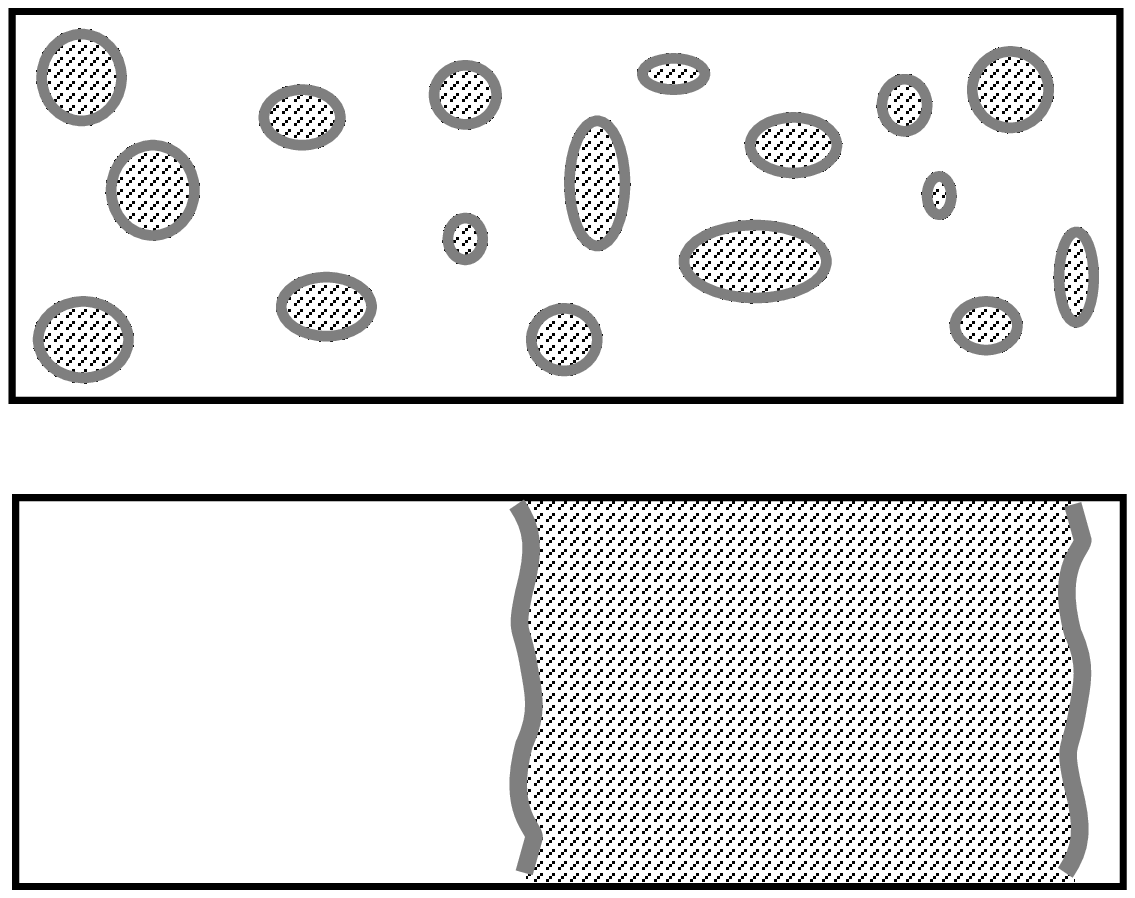}
}
\caption{
(a) Plaquette histogram in the SU(3) gauge theory at the deconfining
transition point obtained on
a $24^2\times36\times4$ lattices\protect\cite{QCDPAX}.
See Fig.~\protect\ref{fig:PlaqHist}(a) for the corresponding
time history.
The two peaks are separately fitted with two Gaussian distributions.
(b) Mix configurations at a first order transition point.
Above: naive configuration with bubbles.
Below: dominant configuration for two phase coexistence on a finite box
with periodic boundary conditions.
}
\label{fig:PlaqHstg}
\end{figure}

\begin{table}[bt]
\caption{Interface tension $\sigma_I/T_c^3$ 
computed using the histogram method for the data of $|\Omega|$.
Large spatial volume limit is taken, except for the results at $N_t=3$
where the spatial lattice volume is $12^3$.
}
\label{tab:interface}
\begin{center}
\begin{tabular}{cllll}
\hline
$N_t$ & standard 
& \multicolumn{2}{c}{Symanzik action\protect\cite{Beinlich97}}
& fixed point  \\
 & action\protect\cite{ourInterface94} 
 & tree-level & MF improved & action\protect\cite{Papa97}\\
\hline
2 & 0.092(4)\protect\cite{Grossmann97}   & & &         \\
3 &            & 0.2434(24) & 0.0158(11) & 0.0307(8)   \\
4 & 0.0295(21) & 0.0152(26) & 0.0152(20) & 0.026(5)    \\
6 & 0.0218(33) &            &            &             \\
\hline
\end{tabular}
\end{center}
\end{table}

Because the deconfining transition is first order for SU(3),
we expect that hot and cold phases can coexist just at the 
transition temperature.
When the transition is also of first order in full QCD,
the value of the surface tension for the interface between two phases 
is important in a study of hadronic bubble formation 
in the cooling process of the quark gluon plasma 
at the early Universe, heavy ion collisions etc..
To gain experience for the study in full QCD, the interface tension has been 
measured in quenched QCD using various actions.

Recent numerical computations of the interface tension are 
based on the ``histogram method''\cite{Binder81}:
As shown in Fig.~\ref{fig:PlaqHist}, the Monte Carlo time history of
an observable that is sensitive to the transition shows 
flip-flops between two phases at the transition point.
Accordingly, when we plot a histogram from the history, 
we obtain two peaks corresponding to the two phases, when the lattice
size is sufficiently large.
As shown in Fig.~\ref{fig:PlaqHstg}(a) for the case of the plaquette
on a $24^2\times36\times4$ lattice, the valley between the two peaks 
is in general higher than that expected from 
an overlap of two distributions corresponding to the two peaks.
This excess is due to the contribution of transition stages (mixed states)
around the flip-flops discussed in Sec.~\ref{sec:ep}. 
Naively, a mixed state would look like bubbles of one phase floating in 
the sea of another phase.
With positive interface tension, however, the dominant contribution will 
be the configurations with minimum interface area.
See Fig.~\ref{fig:PlaqHstg}(b).
Then, because the interface area is approximately fixed from
the lattice geometry, we can calculate the probability to have such a mixed
state as a function of the interface tension.
Comparing the result with the measured probability obtained from the
histogram, we can compute the value of the interface tension\cite{Binder81}.

The actual procedure in lattice QCD is much more complicated.
We perform a finite size scaling analysis taking into account 
various corrections including those from capillary wave excitations 
on the interface\cite{ourInterface94}.
Recent results from various actions are summarized in 
Table~\ref{tab:interface}.
A naive continuum extrapolation using the data for $N_t=4$ and 6
from the standard action gives $\sigma_I/T_c^3 = 0.16(4)$,
which is close to the values obtained with Symanzik improved
actions at $N_t=4$\cite{Beinlich97}, 
but slightly smaller than the value from a fixed point action\cite{Papa97}.

\section{Finite temperature QCD with dynamical quarks}
\label{sec:fullQCD}

Finally, let us study finite temperature QCD with dynamical quarks.
As discussed in Sec.~\ref{sec:MC}, a full QCD simulation requires 
several hundred times more computational time compared with a quenched 
simulation to achieve corresponding accuracy.
Therefore, in many cases, results for full QCD are not yet quite 
quantitative.

In this lecture, we concentrate on the topics of the order of the 
finite temperature transition in full QCD.
With dynamical quarks, the center Z($N_c$) transformation 
(\ref{eq:center}) is no longer a symmetry of the action.
Therefore, although the Polyakov loop is still a good ``indicator''
of the deconfining transition (it sensitively changes its magnitude 
at the transition point), the group Z($N_c$) is no longer a good guide 
to study the nature of the QCD transition.

On the other hand, when quarks are light, we expect that the chiral
symmetry, which is spontaneously broken in low temperatures, is 
recovered when the temperature is sufficiently large.%
\footnote{
Numerical simulations show that, when the quarks are in the fundamental
representation of SU($N_c$), the deconfining transition in the heavy 
quark mass limit smoothly turns into the chiral transition when we 
decrease the quark mass.
}
Using the universality hypothesis, 
the nature of the finite temperature QCD 
transition near the chiral limit 
can be studied by a Ginzburg-Landau effective theory
respecting the chiral symmetry of QCD, 
the effective $\sigma$ model.
From a study of the effective $\sigma$ model at finite 
temperatures\cite{PisarskiWilczek},
the transition in the chiral limit (the chiral transition) is predicted 
to depend quite sensitively on the number of light quark flavors $N_F$.
Let us consider QCD with $N_F$ degenerate light quarks.
For $N_F \geq 3$, a first order transition is predicted from the
sigma model.%
\footnote{We restrict ourselves to the case $N_F \leq 6$.
See Ref.~\citen{ourYKISmanyNf} for the phase structure at $N_F\geq7$.
}
For $N_F=2$, on the other hand, the order of the transition 
is not quite definite in the effective $\sigma$ model; 
a first order transition is predicted when the anomalous axial 
U$\!_A$(1) symmetry 
is effectively restored at the transition temperature, 
while a second order transition is expected otherwise.
However, because the U$\!_A$(1) breaking 
operator is a relevant operator whose coefficient grows 
towards the IR limit under a renormalization group transformation, 
the transition is more likely to be second order\cite{Ukawa95}. 
A non-perturbative study is required to determine the order of the 
transition for $N_F=2$ conclusively. 

In nature, we know six flavors of quarks; u, d, s, c, b, and t. 
The lightest two quarks, u and d, are much lighter 
than the relevant energy scale for thermal processes near 
the critical temperature $T_c \simeq 100$--200 MeV: 
$m_u$, $m_d \ll T_c$. 
On the other hand, the last three quarks, c, b, and t, are 
sufficiently heavy that they are expected to play no appreciable 
roles in thermal processes near $T_c$. In the following, 
the case $N_F=2$ corresponds to the case where the third 
quark s is much heavier than the relevant energy scale; $m_s \gg T_c$, 
while the case $N_F \geq 3$ corresponds to the case $m_s \ll T_c$.
Because $m_s \simeq 150$--200 MeV is just of the same order 
of magnitude as the expected values of $T_c$, 
in order to make a reliable prediction for the real world, 
we have to fine-tune the value of $m_s$ in the 
more realistic case of two light quarks and one heavy quark ($N_F=2+1$). 

\subsection{Chiral transition for $N_F=2$ QCD}
\label{sec:fullQCD_Nf2}

Understanding the nature of the QCD transition for $N_F=2$ is 
an important step toward the clarification of the transition 
in the real world. 
When the transition in the chiral limit (the chiral transition) is 
second order, we expect that 
the transition turns into an analytic crossover at non-zero $m_q$, 
while when the chiral transition is first order, it will remain
to be first order for small $m_q$. 
In order to confirm the expected crossover numerically, 
we have to study the lattice size dependence to see if the 
formation of a singularity 
(e.g.\ the increase of the peak height of a susceptibility with increasing 
the lattice volume) stops on sufficiently large lattices.
However, it is difficult to numerically distinguish between 
a very weak first order transition and a crossover, 
especially at small $m_q$. 

Here, the universality provides us with useful scaling relations 
that can be confronted with numerical results of QCD,
in order to test the nature of the transition: 
It is plausible from an effective $\sigma$ model
that, when the chiral transition 
is of second order, QCD with two flavors
belongs to the same universality class as the three 
dimensional O(4) Heisenberg model\cite{PisarskiWilczek}.
The O(4) model is much simpler than the $\sigma$ model,
and its scaling properties are well studied. 
For example, 
at small external field $h$ near the critical temperature $T_c$ for $h=0$, 
the pseudo-critical temperature $T_{pc}(h)$ and 
the peak height of the magnetic and thermal susceptibilities 
follow $T_{pc}-T_c \sim h^{z_g}$, 
$\chi_m^{\rm max} \sim h^{-z_m}$, and 
$\chi_t^{\rm max} \sim h^{-z_t}$, 
where $z_g = 1/\beta\delta$, $z_m=1-1/\delta$, 
and $z_t=(1-\beta)/\beta\delta$
in terms of the O(4) critical exponents $\beta$ and $\delta$. 
Here the values of $\beta$ and $\delta$ for the O(4) model 
are well established\cite{KanayaKaya}. 
In QCD, we identify $T \sim 6/g^2$, $h \sim m_q$, and 
$M \sim \langle \bar\Psi \Psi\rangle$. 

In lattice QCD, an additional complication should be noted
because no known lattice fermions have 
the full chiral symmetry on finite lattices, 
as discussed in Sec.~\ref{sec:formulation}.
In particular, on a coarse lattice used in a finite temperature simulation,
we sometimes encounter sizable deviations from the scaling behavior
expected in the continuum limit.
Therefore, the appearance of the O(4) scaling is also a useful 
touchstone to test the recovery of the chiral symmetry on 
the lattice when the chiral transition is of second order.

\subsubsection{Results with staggered quarks}
\label{sec:JLQCD}

The O(4) scaling was first tested on the lattice for staggered quarks 
by the Bielefeld group\cite{KarschLaermann}. 
Based on simulations on an $8^3\times4$ lattice at $m_qa=0.02$,
0.0375, and 0.075 using the standard action, they obtained 
$z_g = 0.77(14)$, $z_m = 0.79(4)$, and $z_t = 0.65(7)$, 
where the corresponding O(4) values\cite{KanayaKaya} are 0.537(7), 
0.794(1), and 0.331(7). 
The result for $z_m$ is consistent with the O(4) value 
while other exponents are in disagreement with the O(4) values. 

\begin{figure}[t]
\centerline{
a)\epsfxsize=6.5cm \epsfbox{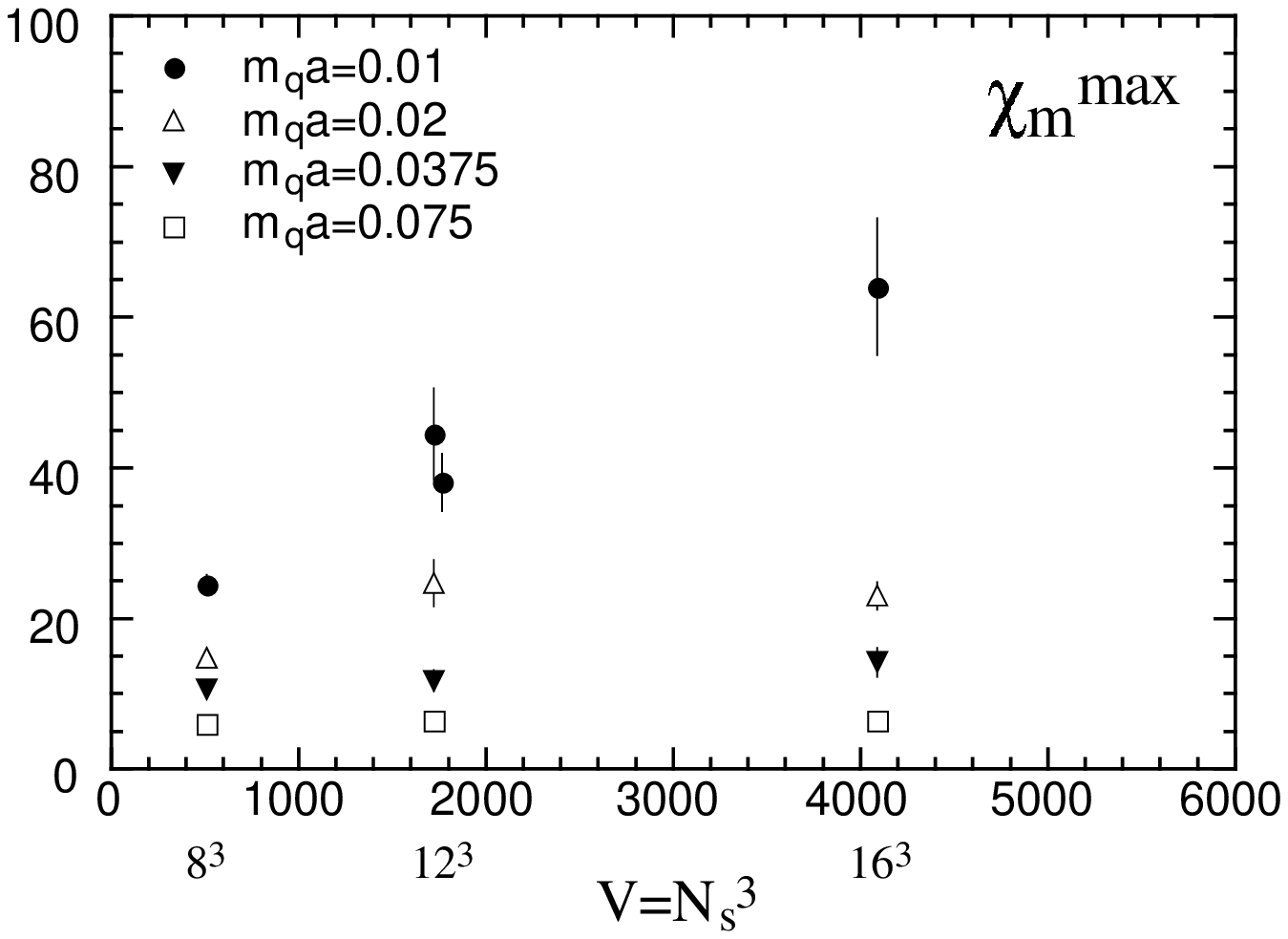}
\makebox[2mm]{}
b)\epsfxsize=6.5cm \epsfbox{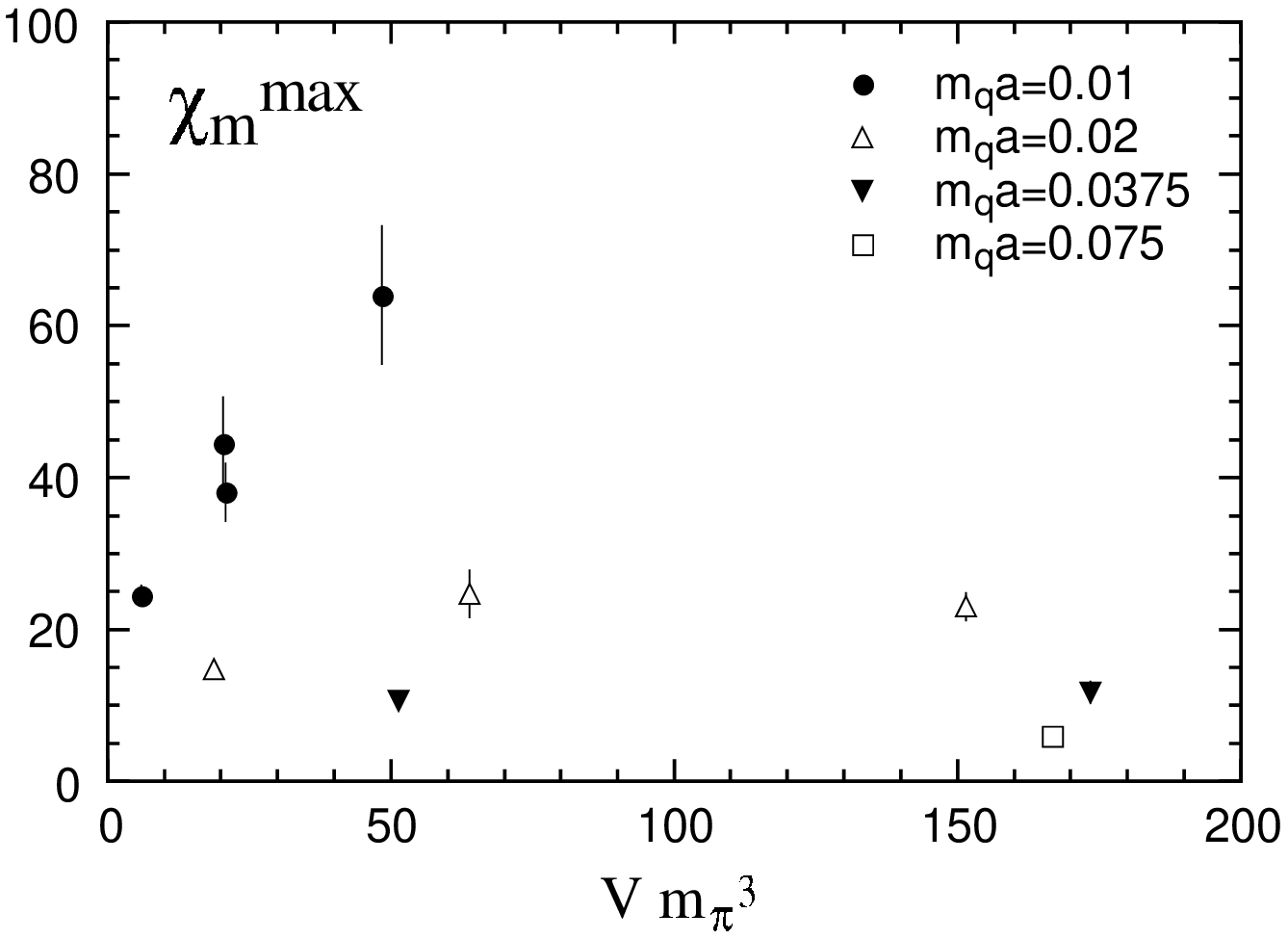}
}
\caption{
(a) Peak height of the magnetic susceptibility 
for $N_F=2$ QCD with staggered quarks 
as a function of the spatial lattice volume $N_s^3$.
(b) The same data as a function of the lattice volume
rescaled by zero-temperature pion correlation length.
}
\label{fig:chim}
\end{figure}

Possible causes of the discrepancy are (i) $m_q$ is not small enough 
to see the critical behavior in the chiral limit, and (ii) 
the spatial lattice volume is not large enough to obtain the 
observables in the thermodynamic limit. 
Two additional caveats are in order for $N_F=2$ staggered quarks: 
(iii) The symmetry in the chiral limit at $a>0$ is O(2) 
instead of O(4). 
Practically, however, the values of the O(2) exponents are almost 
indistinguishable from the O(4) values with the present numerical accuracy.
(iv) The action is not local. 
Therefore, an assumption behind the universality argument 
can be violated so that some non-universal behavior may appear 
\cite{KanayaLat95}.
The correct continuum chiral limit with the O(4) symmetry will
be obtained only when we first take the continuum limit $a \rightarrow 0$ 
and then take the chiral limit. 
In addition to these points, 
we also have to check technical details in the numerical
simulation; the accuracy of the methods to simulate the 
system, such as the the finite step-size error and the dependence
on the convergence criterion for fermion matrix inversion.

A systematic study of the quark mass dependence as well as the lattice
volume dependence is in order.
The JLQCD Collaboration performed a series of simulations on
$8^3\times4$, $12^3\times4$, and $16^3\times4$ lattices at 
$m_qa=0.01$, 0.02, 0.0375, and 0.075 \cite{JLQCDfinT}.
The Bielefeld group also extended their study to larger spatial 
lattices \cite{KarschLaermann2}.
The results obtained are consistent with each other.
It turned out that determination of critical exponents on 
$8^3\times4$ lattices suffers from a sizable finite lattice-size 
effect for $m_qa < 0.0375$.

From the lattice-size dependence of the magnetic susceptibility $\chi_m$
for a fixed value of quark mass, shown in Fig.~\ref{fig:chim}(a),
we find that the transition is a crossover for $m_qa\geq 0.02$; 
the peak height $\chi_m^{\rm max}$ for $m_qa=0.02$
stabilizes on spatial lattices larger than $12^3$.
For $m_qa=0.01$, on the other hand, $\chi_m^{\rm max}$ is increasing
up to the largest spatial lattice of $16^3$. 
If this increase is maintained up to infinite volume, then 
the transition is first order at this quark mass. 
However, no clear indication of a first-order transition 
are found from the lattice volume dependence of Monte Carlo time 
histories and histograms at $m_qa=0.01$.
Furthermore, when the lattice volume is rescaled by the zero-temperature 
pion correlation length,
the lattice volume $16^3$ for $m_qa=0.01$ approximately
corresponds to the volume $12^3$ for $m_qa=0.02$,
where the increase of $\chi_m^{\rm max}$ terminates 
[see Fig.~\ref{fig:chim}(b)].  
Therefore, it is possible that the increase of
$\chi_m^{\rm max}$ for $m_qa=0.01$ is a transient effect.

Assuming that the finite size effect is sufficiently small on the
$16^3\times4$ lattice, we fit the data at the four values of $m_qa$.
We find $z_g=0.64(5)$, $z_m=1.03(9)$, and $z_t=0.82(12)$.
(Removing the data for $m_qa=0.01$ gives slightly smaller
but consistent values with larger errors.)
The results for $z_g$ and $z_m$ deviate sizably 
from the O(2) or O(4) values.
On the other hand,
the identity $z_g+z_m-z_t=1$ expected for a second-order fixed point with
two relevant operators is approximately satisfied.
Thus, 
the exponents are consistent with a second-order transition at $m_q=0$.

In summary, we find that the determination of the nature of the two-flavor
chiral transition with staggered quarks using the standard action 
to involve subtle problems.  
While the data so
far do not contradict a second-order transition at $m_q=0$, the exponents
take quite unexpected values, at least in the range $m_qa \geq 0.01$.  
Evidently further work, possibly on larger
spatial sizes and smaller quark masses, is needed to clarify this
important problem.

\subsubsection{Results with Wilson quarks}
\label{sec:TsukubaNf2}

Let us now study the issue using Wilson quarks.
It turned out that Wilson quarks in the standard action lead to 
several unexpected phenomena on lattices with $N_t=4$ and 6:
On these lattices, the transition becomes once very sharp when 
$m_q$ is increased from the chiral limit \cite{MILC46,ourStandard96},
contrary to the expectation in the continuum limit that 
the chiral transition becomes weaker with larger $m_q$.
Together with other strange behavior of physical quantities 
near the transition point, this phenomenon is identified as an effect of 
lattice artifacts \cite{ourStandard96}. 

\begin{figure}[t]
\begin{center}\leavevmode
\epsfxsize=14cm \epsfbox{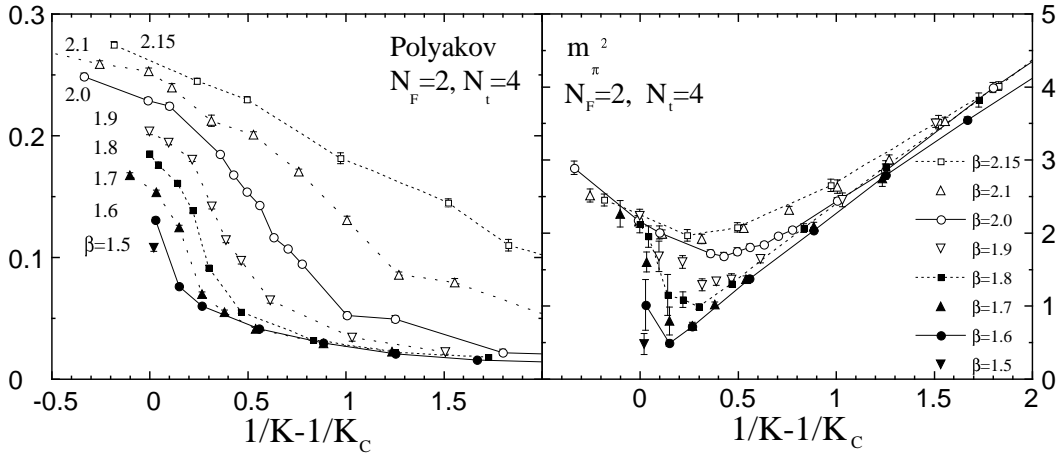}
\end{center}
\vspace{-0.2cm}
\caption{
The Polyakov loop and the pion screening mass 
obtained with Wilson quarks with a RG improved action
\protect\cite{ourPRL97}.
In these figures, the conventional notation $\beta \equiv 6/g^2$ 
is used.
Larger $\beta$ corresponds to a higher temperature.
The horizontal axis $1/K-1/K_C$ is proportional to $m_q a$.
}
\label{fig:iF2T4PMpi}
\end{figure}

Therefore, the Tsukuba group applied an improved action \cite{ourPRL97}. 
Their action is the RG improved gauge action (\ref{eq:Iwasaki}) 
with $c_1 = - 0.331$, 
coupled with the standard Wilson quark action.
Although the quark part is not improved, 
the lattice artifacts observed with the standard action
are shown to be well removed\cite{ourPRL97,ImFQcdTsukuba}. 
They also find that the physical quantities are quite smooth around the
transition point at $m_q>0$, as shown in Fig.~\ref{fig:iF2T4PMpi}.
The straight line envelop of $m_\pi^2$ at finite temperature ($N_t=4$) 
shown in Fig.~\ref{fig:iF2T4PMpi}(b) 
agrees with $m_\pi^2$ obtained at low temperature ($N_t=8$), 
and corresponds to the PCAC relation $m_\pi^2 \propto m_q$ expected in the 
low-temperature phase.
The smoothness of the physical observables strongly suggests that 
the transition is a crossover at $m_q > 0$. 

Concerning the nature of the transition in the chiral limit,
the transition becomes monotonically weaker with increasing $6/g^2$
(see Fig.~\ref{fig:iF2T4PMpi}).
Because the transition point shifts to larger $6/g^2$ at larger $m_q$, 
increasing $6/g^2$ corresponds to increasing $m_q$ for the transition.
In the chiral limit $m_\pi^2$ decreases monotonically
to zero as the temperature is decreased from above, 
towards the chiral transition point. 
At the transition temperature for finite $m_q$, $m_\pi^2$ 
also shows a similar monotonic decrease with decreasing $m_q$.
These results suggest that the chiral transition is continuous. 

\begin{figure}[t]
\centerline{
a)\epsfxsize=6.3cm\epsfbox{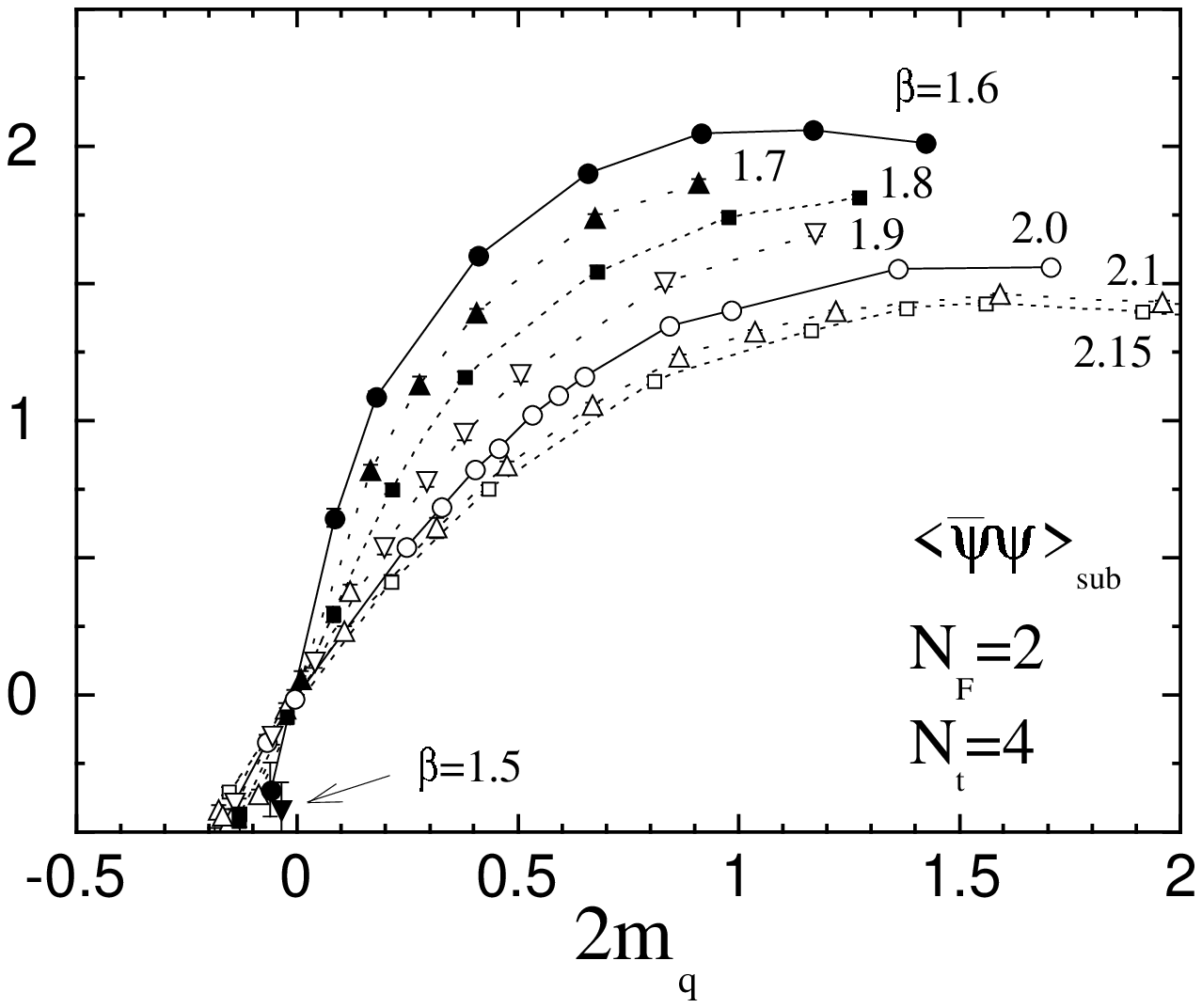}
\makebox[1mm]{}
b)\epsfxsize=6.5cm\epsfbox{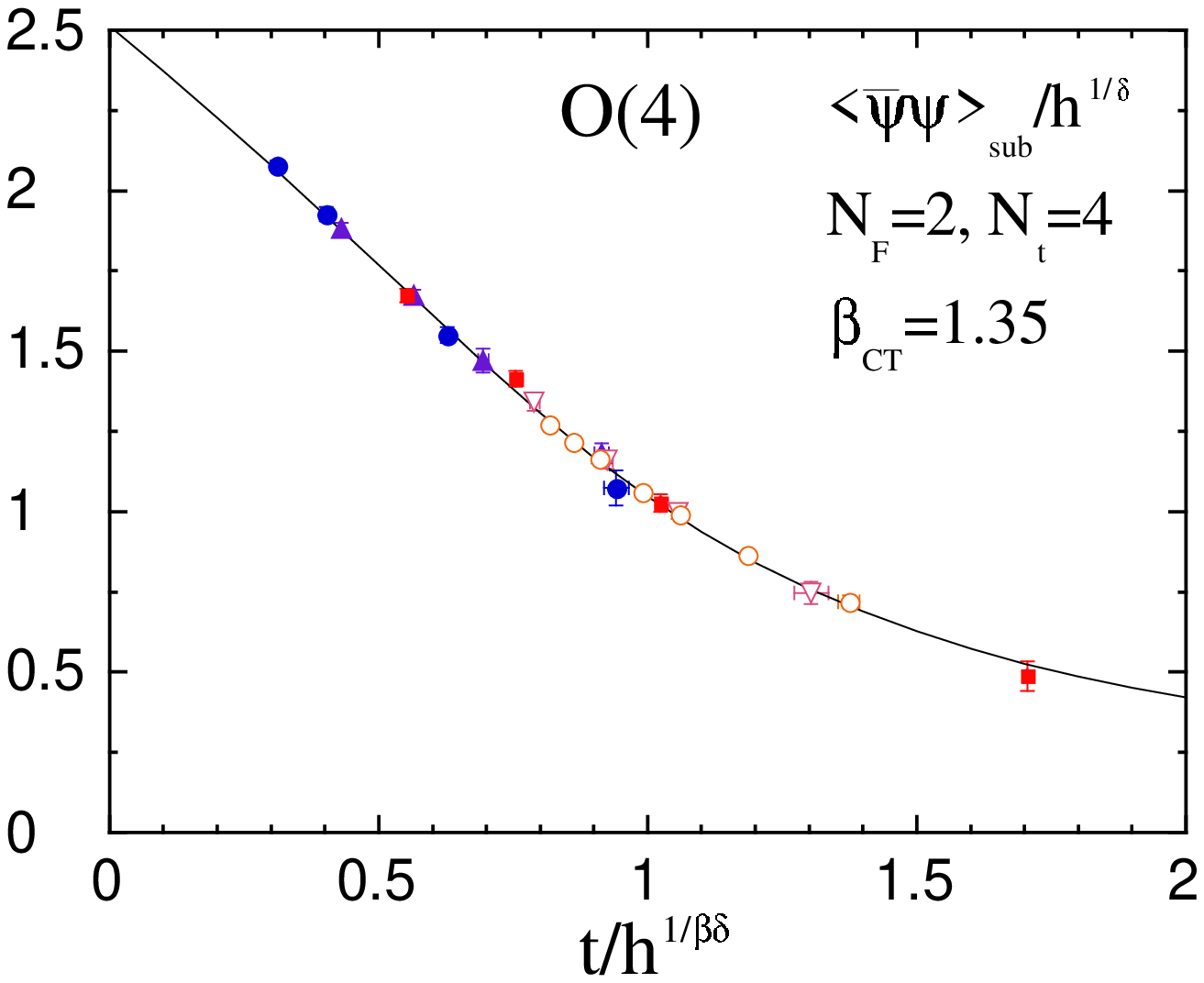}
}
\caption{(a) 
Chiral condensate as a function of $h=2m_qa$ 
for Wilson quarks with a RG improved action\protect\cite{ourPRL97}.
(b)
Best fit for the scaling function with O(4) exponents 
with $\chi^2/df = 0.61$, 
using the identification $h=2m_qa$ and $t=\beta-\beta_{ct}$, 
where $\beta \equiv 6/g^2$ and 
$\beta_{CT}$ is the value of $6/g^2$ at the chiral transition point.
Do not confuse with the critical exponent $\beta$ 
appearing in the combination $t/h^{1/\beta\delta}$.
The plot contains all data of (a) 
within the range $0 < 2m_q a < 0.8$ and $6/g^2 \leq 2.0$. 
Solid curve represents the scaling function obtained in an O(4) 
spin model. 
}
\label{fig:pbp}
\end{figure}

For a more decisive test about the nature of the transition, 
a scaling study is required.
From the universality argument, the magnetization $M$ in a spin model 
can be described by a single scaling function near a second order 
transition point: 
\begin{equation}
M / h^{1/\delta} = f(t/h^{1/\beta\delta}),
\end{equation}
where $h$ is the external magnetic field 
and $t=[T-T_c]/T_c$ the reduced temperature. 
When the QCD transition is of second order in the chiral limit, 
the chiral condensate should satisfy this scaling relation 
with O(4) exponents $1/\beta\delta = 0.537(7)$ and $1/\delta = 0.2061(9)$ 
\cite{KanayaKaya} 
and also with the O(4) scaling function $f(x)$.

The magnetization $M$ is identified with the chiral condensate in QCD.
For Wilson quarks, 
the naive definition of $\langle \bar{\psi} \psi \rangle$
for the chiral condensate 
is not adequate because the chiral symmetry is explicitly 
broken due to the Wilson term. 
A proper subtraction and a renormalization are required to obtain 
the correct continuum limit. 
A properly subtracted 
$\langle \bar{\psi} \psi \rangle$ can be defined via an axial 
Ward identity \cite{Bochicchio}:
\begin{equation}
\langle \bar{\psi} \psi \rangle_{\rm sub} 
= 2 m_q a Z \sum_x \langle \pi(x) \pi(0) \rangle
\label{eq:PBPsub}
\end{equation}
where $Z$ is the renormalization coefficient. 
For our purposes, it is enough to use the tree value,
$Z=(2K)^2$. 
When the chiral symmetry is spontaneously broken, a singulariry in
the pion propagator cancels the factor $m_q$ in the r.h.s.\ of 
(\ref{eq:PBPsub}),
giving a finite value for $\langle \bar{\psi} \psi \rangle_{\rm sub}$ 
in the chiral limit.

The results of $M = \langle \bar{\psi} \psi \rangle_{\rm sub}$ 
around the deconfining transition/crossover for $N_t=4$ 
is shown in Fig.~\ref{fig:pbp}(a).
Fig.~\ref{fig:pbp}(b) shows the result from a fit of $M$ 
to the scaling function obtained for an O(4) model \cite{Toussaint},
by adjusting the chiral transition point $\beta_{ct}$ and the scales 
for $t$ and $h$, with the exponents fixed to the O(4) values.
The scaling ansatz works remarkably well with the O(4) exponents. 
A recent study shows that the situation holds also when data at $t \leq 0$ 
are included \cite{ImFQcdTsukuba}.
On the other hand, a change of the exponents quickly makes the fit worse:
For example, fixing the exponents to the MF values, suggested by Koci\'c and
Kogut as a possibility for two-flavor QCD\cite{KocicKogut},
the data no longer falls on the MF scaling function. 

The success of this scaling test with the O(4) exponents 
suggests strongly that the chiral transition is of second order 
in the continuum limit. 
It also indicates that 
the chiral violation due to the Wilson fermion action is sufficiently 
small with this improved action, 
for the values of $m_q$ and $6/g^2$ studied here. 

\subsection{Influence of the strange quark}
\label{sec:strange}

\begin{figure}[tb]
\centerline{
a)\epsfxsize=6.5cm \epsfbox{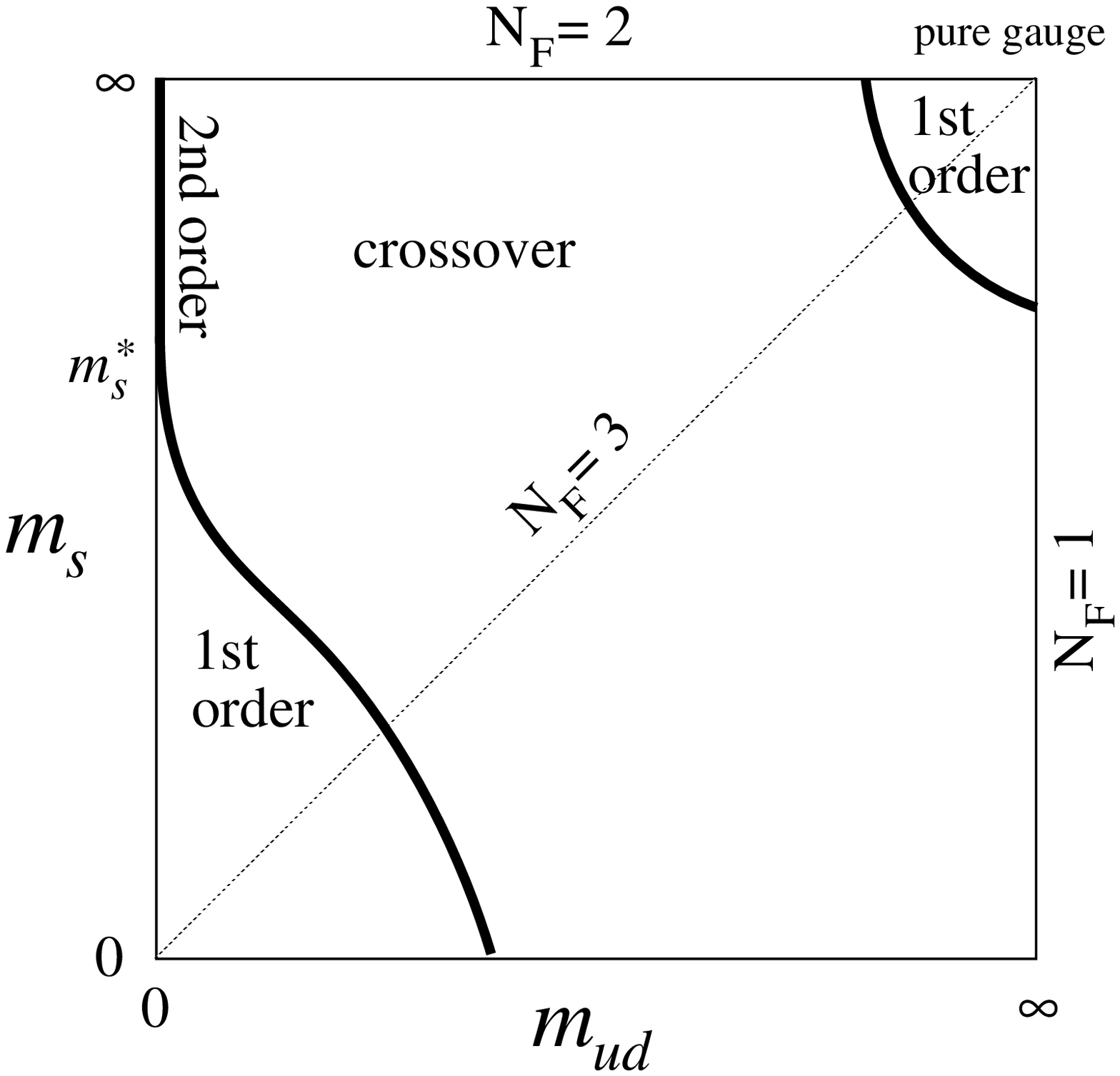}
\makebox[2mm]{}
b)\epsfysize=6.3cm \epsfbox{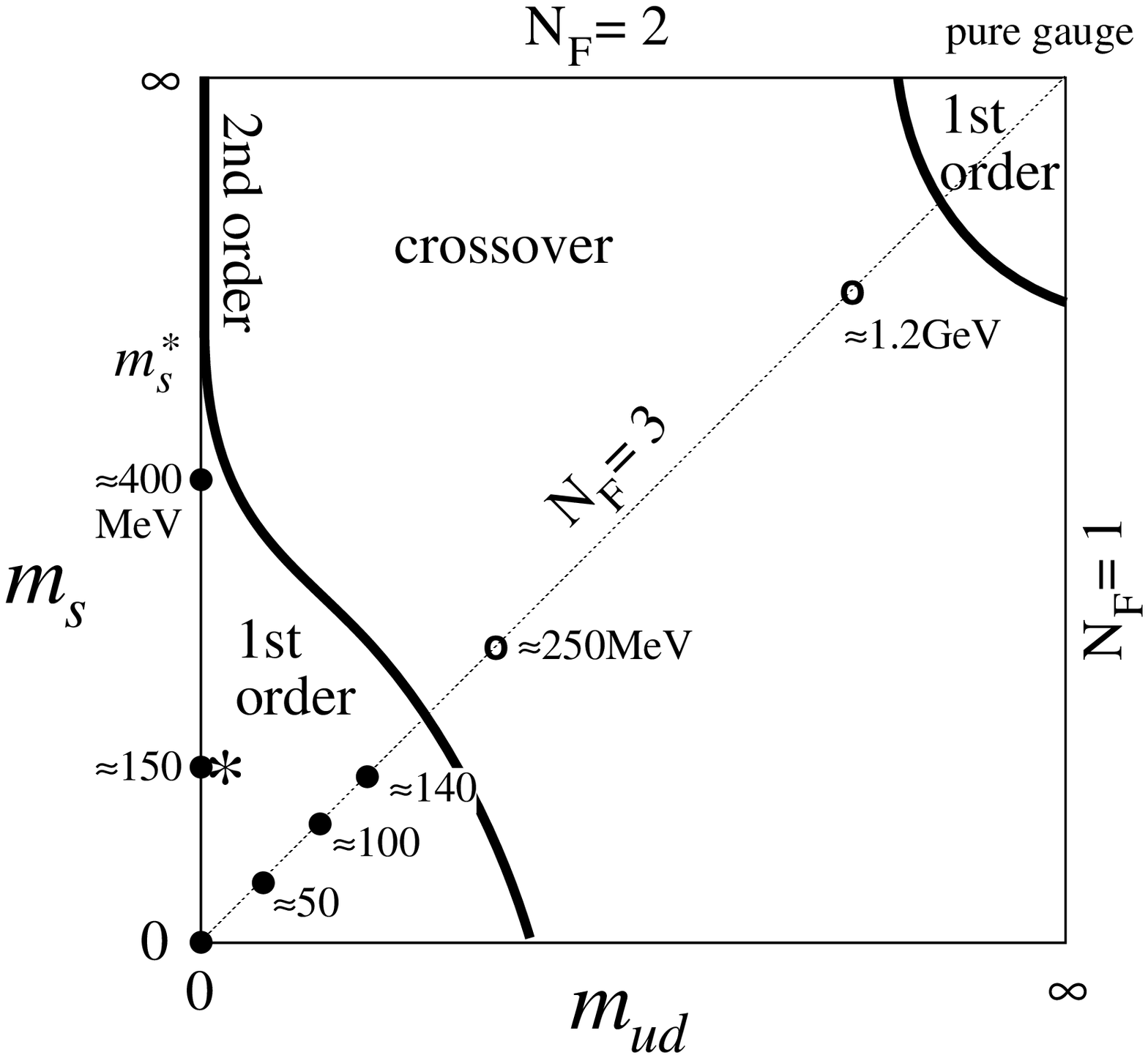}
}
\caption{
(a) Map of expected nature of the QCD transition for $N_F=2+1$ QCD 
as a function of the u and d quark mass $m_{ud}$ 
and the s quark mass $m_s$. 
(b) The same figure with the results from 
Wilson quarks using the standard action \protect\cite{ourStandard96}. 
First order signals are observed at the points marked by 
filled circles, while no clear two-state signals are found 
at the points represented by the open circles. 
The values of quark mass in physical units are computed 
using $a$ 
determined by $m_\rho(T\!=\!0)=770$ MeV. 
The real world corresponds to the point marked by the star. 
}
\label{fig:F2+1}
\vspace{-0.3cm}
\end{figure}

In order to study the nature of the transition in the real world, 
we should study the influence of the s quark. 
Our expectation about the nature of the finite temperature 
transition as a function of quark masses is summarized in
Fig.~\ref{fig:F2+1}(a), 
neglecting the mass difference among u and d quarks ($N_F=2+1$).
The limit $m_s = \infty$ corresponds to the case $N_F=2$ 
discussed in Sect.~\ref{sec:fullQCD_Nf2} where we found 
second order transition at $m_{ud}=0$. 
For $m_{ud}=m_s$ ($N_F=3$), the transition is of first order 
in the chiral limit. 
Therefore, on the axis $m_{ud}=0$, we have a tricritical 
point $m_s^*$ 
where the second order transition at large $m_s$ 
turns into first order \cite{PisarskiWilczek}. 
For $m_s < m_s^*$, the second order edge of the first order 
transition region is suggested\cite{Rajagopal} to deviate 
from the vertical axis 
according to $m_{ud} \propto (m_s^{*} - m_s)^{5/2}$.
Our main goal of investigations with the s quark is to determine
the position of the physical point in this map. 

With staggered quarks, 
Brown {\it et al.} \cite{Columbia} found, 
for the degenerate $N_F=3$ case ($m_{ud} = m_s \equiv m_q$), 
a first order signal at $m_q a = 0.025$, $\beta = 5.132$. 
For $N_F=2+1$, they obtained a time history 
suggesting a crossover 
for $m_{ud}a=0.025$, $m_s a=0.1$ at $\beta=5.171$. 
Their study of hadron spectrum at this simulation point 
on a zero-temperature lattice 
leads to a $m_K/m_\rho$ smaller than the experimental value, 
suggesting that this $m_s$ is smaller than the physical value. 
At the same time, their large $m_\pi / m_\rho$ suggests that 
their $m_{ud}$ is larger than the physical value. 
This implies that the physical point is located in the crossover region 
unless the second order transition line, 
which has a sharp $m_{ud}$ dependence 
near $m_s^*$ (cf.\ Fig.~\ref{fig:F2+1}(a)), 
crosses between the physical point and the simulation point. 
In order to obtain a more decisive conclusion, a study to
systematically investigate a wider region of the parameter 
space in Fig.~\ref{fig:F2+1}(a) is required.

The Tsukuba group studied the issue with Wilson quarks using 
the standard action \cite{ourStandard96}.
They found first order signals for $m_q \simm{<} 140$ MeV,
while no clear two state signals were observed 
for $m_q \simm{>} 250$ MeV, 
where the physical s quark mass, giving $m_\phi = 1.02$ GeV, 
is about 150 MeV in their normalization of $m_q$.
(The scale was fixed by $m_\rho$ at zero tmeperature.)
For $N_F=2+1$, first order signals are observed for 
$m_{ud} \sim 0$ at both $m_s \sim 150$ and 400 MeV. 
A study of zero-temperature hadron spectroscopy for $N_F=2+1$ 
shows that $m_\phi \sim 1.03(5)$ GeV 
at the simulation point $m_s \sim 150$ MeV,
verifying that this simulation point is very close to 
the physical point. 
The results are summarized in Fig.~\ref{fig:F2+1}(b).
The physical point is located in the first order region. 

Although both staggered and Wilson simulations give a phase 
structure qualitatively consistent with 
Fig.~\ref{fig:F2+1}(a), 
Wilson quarks tend to give larger values for critical 
quark masses (measured by $m_\phi / m_\rho$ etc.) 
than those with staggered quarks. 
This leads to the difference in the conclusions 
about the location of the physical point in Fig.~\ref{fig:F2+1}(a). 
On the other hand, both of these studies 
discuss sizable deviation of several physical observables 
from the experimental values, meaning that 
the deviation from the continuum limit is 
large at $N_t=4$ where these simulations are done. 
We should certainly carry out a calculation at larger $N_t$ 
or with an improved action 
in order to draw a definite conclusion about the nature of 
the QCD transition in the real world.

\section{Summary}
\label{sec:summary}

With the present power of computers, we can perform reliable extrapolations
to the continuum limit for several physical quantities 
in the quenched approximation of QCD.  
Precise values of the transition temperature, latent heat, 
interface tension, etc.\ obtained from detailed finite size scaling
analyses are discussed in the literature.
For full QCD simulations, however, several hundred times more computer 
time is required compared to quenched simulations.
This corresponds to about 5--10 years difference in the development
of computer speed.
While a few projects to construct a computer with such a speed have been
proposed, we are not simply waiting for new computers.
Many theoretical developments, especially the progress in improving
lattice actions,
open us the possibility to begin realistic simulations
of full QCD on present supercomputers.
Applications of improved actions to full QCD at finite temperatures have 
just begun.\cite{ourPRL97,ImFQcdTsukuba,ImFQcdMILC_FT,ImFQcdBielefeld}

\section*{ACKNOWLEDGEMENTS}
\vspace{-3mm}
I would like to thank the participants of YKIS'97 for 
valuable comments.
I am also grateful to my colleagues, R.\ Burkhalter, S.\ Ejiri, 
Y.\ Iwasaki, T.\ Kaneko, S.\ Kaya, H.\ Shanahan, A.\ Ukawa and 
T.\ Yoshi\'e for their support and useful discussions.
This work is in part supported by the Grants-in-Aid
of Ministry of Education (Nos.~08NP0101 and 09304029),
and also in part by the Supercomputer Project 
of High Energy Accelerator Research Organization (KEK).

\end{document}